\documentclass[prd,twocolumn,superscriptaddress,altaffilletter,showpacs,nofootinbib]{revtex4}

\usepackage{amssymb}
\usepackage[dvips]{graphicx}
\usepackage{amsmath}

\newcommand{\be}{\begin{equation}}
\newcommand{\ee}{\end{equation}}
\newcommand{\bea}{\begin{eqnarray}}
\newcommand{\eea}{\end{eqnarray}}

\newcommand{\der}{\partial}
\newcommand{\vphi}{\varphi}

\begin{document}

\title{On local scale invariance and the questionable theoretical basis of the conformal transformations' issue}

\author{Israel Quiros}\email{iquiros@fisica.ugto.mx}\affiliation{Departamento de Ingenier\'ia Civil, Divisi\'on de Ingenier\'ia, Universidad de Guanajuato, Gto., M\'exico.}

\author{Roberto De Arcia}\email{robertodearcia@gmail.com}\affiliation{Departamento de Ingenier\'ia Civil, Divisi\'on de Ingenier\'ia, Universidad de Guanajuato, Gto., M\'exico.}

\date{\today}

\begin{abstract}Here we follow the mainstream of thinking about physical equivalence of different representations of a theory, regarded as the consequence of invariance of the laws of physics -- represented by an action principle and the derived motion equations -- under given transformations; be it coordinate, gauge or conformal transformations. Accordingly the conformal transformations' issue is discussed by invoking the assumed invariance of the laws of physics -- in particular the laws of gravity -- under conformal transformations of the metric. It is shown that Brans-Dicke and scalar-tensor theories are not well-suited to address physical equivalence of the conformal frames since the corresponding laws of gravity are not invariant under the conformal transformations or Weyl rescalings. The search for conformal symmetry leads us to explore the physical consequences of Weyl-invariant theories of gravity, that represent a natural arena where to discuss on physical equivalence of the conformally related representations. We show that conformal invariance of the action of a (supposedly conformal invariant) theory and of the derived motion equations is not enough to ensure actual Weyl invariance. It is required, also, that the underlying geometrical structure of the background spacetime be, at least, Weyl-integrable. Otherwise, if assume (as usual) spacetimes of (pseudo)Riemannian geometrical structure, the resulting -- apparently conformal invariant -- theory is anomalous in that, only massless matter fields can be consistently coupled. Gauge freedom, a distinctive feature of actually Weyl-invariant theories of gravity, leads to very unusual consequences. In a conformal invariant gravity theory over Weyl-integrable spaces, when working within the cosmological setting, an interesting question would not be, for instance, whether the expansion of the universe is accelerating or not, but, which one of the infinitely many physically equivalent conformal universes is the one we live in. It happens, also, that static spherically symmetric black holes are physically equivalent to singularity-free spherically symmetric wormholes. Such apparently senseless consequences in standard scalar-tensor theories, are perfectly allowed in anomaly-free conformal invariant theories of gravity.\end{abstract}

\pacs{04.20.Cv, 04.50.Kd, 11.15.-q, 11.30.-j, 98.80.Cq}

\maketitle

%%%%%%%%%%%%%%%%%%%%%%%%%%%%%%%%%%%%%%%%

\section{Introduction}\label{intro}  

The so called conformal frames' issue is, perhaps, one of the oldest unsolved problems within the framework of the scalar-tensor theories of gravity \cite{dicke-1962, faraoni-book, faraoni_rev_1997, faraoni_ijtp_1999, faraoni_prd_2007, sarkar_mpla_2007,  sotiriou_etall_ijmpd_2008, deruelle, quiros_grg_2013}. The problem may be stated in the following way: Under conformal transformations of the metric the given scalar-tensor theory (STT) may be formulated in a -- in principle infinite -- set of mathematically equivalent field variables, called as conformal frames. Among these the Jordan frame (JF) and the Einstein's frame (EF) are the most outstanding. The following related questions are the core of the ``conformal transformation's issue''.

\begin{enumerate} 

\item Are the different conformal frames not only mathematically equivalent but, also, physically equivalent?

\item If the answer to the former question were negative, then: which one of the conformal frames is the physical one, i. e., the one in terms of whose field variables to interpret the physical consequences of the theory? 

\end{enumerate} 

The controversy originates from the lack of consensus among different researchers -- also among the different points of view of the same researcher along his(her) research history -- regarding their answer to the above questions. There are even very clever classifications of the different works -- of different authors and of the same author -- on this issue \cite{faraoni_rev_1997}. That the controversy has not been resolved yet is clear from the amount of yearly work on the issue where there is no agreement on the correct answer to these questions \cite{saal_cqg_2016, indios_prd_2018, ct-1, ct(fresh-view)-3, paliatha-2, steinwachs, ct(quant-equiv)-4, ct(quant-equiv)-5, ct(quant-equiv)-6, ct(inequiv)-7, ct(inequiv)-8, ct(inequiv)-9}. 

In this paper we shall discuss on the conformal frames' issue from the classical point of view exclusively.\footnote{Our understanding of what a STT entails and why it is different from GR with a scalar field as a matter source of the Einstein's equations, may be correct only if ignore the quantum effects of matter. When these effects are considered, even if we start with GR with a scalar field among the matter degrees of freedom, the quantum interactions of matter may induce a non-minimal coupling of the scalar field with the curvature \cite{callan_ann_phys_1970}, so that we end up with a STT. We have included a brief account of the demonstration given in Ref. \cite{callan_ann_phys_1970}, in the Appendix \ref{sect-callan}.} For a related discussion based on quantum arguments we recommend Refs. \cite{steinwachs, ct(quant-equiv)-4, ct(quant-equiv)-5, ct(quant-equiv)-6, ct(inequiv)-7, ct(inequiv)-8} and references therein. 

A conformal transformation of the metric:

\bea g_{\mu\nu}\rightarrow\Omega^{-2}g_{\mu\nu}\;\left(g^{\mu\nu}\rightarrow\Omega^2g^{\mu\nu}\right),\;\sqrt{|g|}\rightarrow\Omega^{-4}\sqrt{|g|},\label{conf-t}\eea where $\Omega^2=\Omega^2(x)$ is the (non-vanishing) conformal factor, is a point-dependent rescaling of the metric that, besides angles, preserves the causal structure of the spacetime. Here we underline that the above conformal transformation of the metric is not to be confounded with conformal transformations implying simultaneous coordinate rescalings, which are properly diffeomorphisms. In other words, we consider conformal transformations of the metric that leave the spacetime coordinates untransformed. 

While the JF and EF representations of the BD theory -- as well as any other two conformally related frames -- are in a relationship of mathematical equivalence through \eqref{conf-t}, their physical equivalence may be, at least, questionable. First of all we must agree on what to regard as ``physical equivalence'' of the different conformal frames since, otherwise, the discussion is meaningless. The following related arguments are found in the bibliography. In Ref. \cite{dicke-1962}, for instance, the physical equivalence of JF and EF formulations of BD theory is assumed by allowing the units of time, length, mass and the derived quantities to scale with appropriate powers of the conformal factor $\Omega$ (see, also, Ref. \cite{faraoni_prd_2007}). According to the argument given in these references, physics must be invariant under the choice of the units, i. e., under the rescaling of the units of length, time and mass. The logic consequence is that, since physics is invariant under a change of units, it is invariant under a conformal transformation, provided that the units of length, time, and mass are scaled \cite{faraoni_prd_2007}. In this regard such concepts like ``EF with running units'' is encountered. The main idea behind this latter concept is that, what really matters when measurements are concerned, is the ratio of the quantity being measured, for instance the mass $m_p$ of a given particle of the standard model of particles (SMP), to the unit of measurement (say, the energy $m_A$ of some emission line of some atom): $m_p/m_A$, and since the ratio is a dimensionless quantity, it is not transformed by the conformal transformations of the metric \eqref{conf-t}, so that the measured value is the same in the JF and in the EF (or in any of the conformally related frames). No matter how ``natural'' such kind of argument could seem, the fact is that conformal transformations of the metric are about point-dependent rescalings of the metric tensor with the spacetime coincidences, i. e., the coordinates, held fixed. In other words, the conformal transformations do not affect the measurements. Hence the above argument is a redundancy. 

It is even more confusing to understand how, according to the above reasoning line, the conformal transformation between the JF and the EF of the BD theory can be reconciled with the assumed conformal invariance of the physical laws. Recall that the gravitational laws in the JF and in the EF of BD theory are expressed through the action principle and the derived motion equations (see section \ref{sect-bd-jf-ef} below). The problem is that, since BD theory is not invariant under the conformal transformations, the action and the derived motion equations are different in the different conformal frames. This means that Brans-Dicke theory (the same is true for any other STT) does not embody the assumed conformal invariance of the physical laws.

In this regard, one should not be confused by the argument frequently found in the bibliography that the gravitational part of the BD action: $$S_\text{BD}=\int d^4x\sqrt{|g|}\,e^\vphi\left[R-\omega(\der\vphi)^2\right],$$ where we have rescaled the BD field, $\vphi\rightarrow\ln\phi$, and the subindex ``BD'' in the coupling constant $\omega$, has been omitted, is invariant under a conformal transformation of the metric \eqref{conf-t}, plus a transformation of the coupling constant, $$\omega\rightarrow\omega-2\der_\vphi\ln\Omega\left(1-\der_\vphi\ln\Omega\right)(2\omega+3).$$ One should notice first that, actually, the BD action above is form-invariant under the aforementioned transformations. However, these imply that in general a constant value of the coupling constant in the Jordan frame is transformed into a function of $\vphi$ in the conformal frame, although there are cases when, $\der_\vphi\ln\Omega=\alpha$, is a constant, where one constant value of the coupling constant is mapped into a different constant value in the conformal frame. This, in turn, has implications for the measured value of the gravitational constant (the scalar field is determined by appropriate cosmological boundary conditions given far from the system of interest), $$8\pi G_\text{eff}=\frac{1}{\phi}\left[\frac{4+2\omega}{3+2\omega}\right],$$ so that one has actually two different theories: BD theory with different values -- even different behaviors -- of the measured gravitational coupling $G_\text{eff}$.

%---------------------------------------------------------------------------------------------

\subsection{Physical equivalence according to the mainstream of thinking}\label{subsect-intro}

When one thinks about physical equivalence one of the first examples that comes to one's mind is the theory of general relativity. The physical equivalence of the different coordinate frames in which the GR laws -- expressed through the action principle and the derived equations of motion -- can be formulated, is sustained by the invariance of these laws under general coordinate transformations. This leads naturally to the existence of a set of measurable quantities: the invariants of the geometry such as the line element, the curvature scalar and other quantities that are not transformed by the general coordinate transformations. Another example can be the gauge theories, where the gauge symmetry warrants that the theory can be formulated in a set of infinitely many physically equivalent gauges. In this case the quantities that have the physical meaning, i. e., those that are connected with measurable quantities, are gauge-invariant. As before, the guiding principle that supports the physical equivalence of the different gauges is the underlying symmetry. Take as a very simple example the electromagnetic gauge theory of a Fermion field $\psi(x)$, that is given by the following Lagrangian:

\bea {\cal L}_\text{gauge}=\bar\psi(x)\left(iD-m\right)\psi(x)-\frac{1}{4}F_{\mu\nu}F^{\mu\nu},\label{gauge-lag}\eea where the gauge derivative $D\equiv\gamma^\mu(\der_\mu-ig A_\mu)$ ($\gamma^\mu$ are the Dirac gamma-matrices while $A_\mu$ are the electromagnetic potentials) and $F_{\mu\nu}\equiv\der_\nu A_\mu-\der_\mu A_\nu$. The above Lagrangian is invariant under the following gauge transformations: $$\psi(x)\rightarrow e^{i\alpha(x)}\psi,\;\;\bar\psi(x)\rightarrow e^{-i\alpha(x)}\bar\psi,\;\;A_\mu\rightarrow A_\mu+\frac{1}{g}\der_\mu\alpha.$$ Quantities that are invariant under the above transformations, such as, for instance those $\propto \bar\psi\psi$, or $\propto F_{\mu\nu}F^{\mu\nu}$, and the related quantities, are the ones that have the physical meaning. The above procedure can be straightforwardly generalized to a collection of Fermion fields and of gauge fields in the electroweak (EW) theory, for instance.

By analogy, one may expect that physical equivalence of the conformal frames should be linked with conformal invariance of the laws of physics, in particular, of the gravitational laws \cite{dicke-1962, faraoni_prd_2007}. Actually, following the spirit of the above examples: coordinate invariance of the laws of gravity in GR and gauge invariance of the laws of electromagnetism, one should require the action and the field equations of the theory -- representing the physical laws -- to be invariant under \eqref{conf-t}. Then one may search for quantities that are not transformed by the conformal transformations of the metric, and regard them as the measurable quantities of the theory. This is the natural way in which one may think about invariance of the physical laws under conformal transformations and this is, precisely, the main thesis of the present work: Conformal invariance of the physical laws -- expressed through the action principle and the derived motion equations -- is the necessary requirement for physical equivalence of the different conformal frames in which a given STT may be formulated. This leads us to the investigation of local scale invariance or Weyl-invariance within the framework of the gravitational theories.

Local scale invariance is one of the symmetries that has played an important role in the building of the unified interactions. It is required for the renormalization procedure to work appropriately at very short distances \cite{smolin_npb_1979, zee_prd_1981, cheng_prl_1988, goldberger_prl_2008}. In what regards gravitation theories local scale invariance, or Weyl-invariance as it is better known, has been also investigated from different perspectives \cite{smolin_npb_1979, zee_prd_1981, cheng_prl_1988, goldberger_prl_2008, sc_inv_indios, sc_inv_indios_1, sc_inv_shapo, sc_inv_percacci, sc_inv_prester, sc_inv_quiros, sc_inv_padilla, sc_inv_bars, sc_inv_bars_1, sc_inv_bars_2, sc_inv_carrasco, sc_inv_quiros_1, sc_inv_jackiw, sc_inv_alpha, sc_inv_alpha_2, sc_inv_alpha_3, sc_inv_alpha_4, salvio-1, sc_inv_ghorbani, sc_inv_farz, sc_inv_vanzo, sc_inv_alvarez, alvarez-jcap-2015, sc_inv_khoze, sc_inv_karananas, sc_inv_tambalo, sc_inv_kannike, sc_inv_ferreira, sc_inv_maeder, sc_inv_myung}. In Ref. \cite{weyl} the first serious attempt to create a scale-invariant theory of gravity (and of electromagnetism) was made. Due to an unobserved broadening of the atomic spectral lines this attempt had a very short history \cite{sc_inv_scholz_1, sc_inv_scholz_2, sc_inv_perlick, sc_inv_novello}. A scale-invariant extension of general relativity based on Weyl's geometry is explored in Ref. \cite{smolin_npb_1979}. If the theory contains a Higgs phase then, at large distances this phase reduces to GR. In Ref. \cite{zee_prd_1981} it has been shown that gravity may arise as consequence of dynamical symmetry breaking in a scale -- also gauge -- invariant world. A quantum field theory of EW and gravitational interactions with local scale invariance and local $SU(2)\times U(1)$ gauge invariance is proposed in Ref. \cite{cheng_prl_1988}. The requirement of local scale invariance leads to the existence of Weyl's vector meson which absorbs the Higgs particle remaining in the SMP.

In general any theory of gravity can be made Weyl-invariant by introducing a dilaton. In Ref. \cite{sc_inv_percacci} it is shown how to construct renormalization group equations for such kind of theories, while in Ref. \cite{sc_inv_odintsov, sc_inv_odintsov_1, sc_inv_odintsov_2, sc_inv_odintsov_3} it has been shown that scale invariance is very much related with the effect of asymptotic conformal invariance, where quantum field theory predicts that theory becomes effectively conformal invariant. In Ref. \cite{sc_inv_padilla} the authors present the most general actions of a single scalar field and two scalar fields coupled to gravity, consistent with second order field equations in four dimensions (4D), possessing local scale invariance. It has been shown that Weyl-invariant dilaton gravity provides a description of black holes without classical spacetime singularities \cite{sc_inv_prester}. The singularities appear due to ill-behavior of gauge fixing conditions, one example being the gauge in which the theory is classically equivalent to GR. In Refs. \cite{sc_inv_bars, sc_inv_bars_1, sc_inv_bars_2} it is shown how to lift a generic non-scale invariant action in Einstein frame into a Weyl-invariant theory and a new general form for Lagrangians consistent with Weyl symmetry is presented. Advantages of such a conformally invariant formulation of particle physics and gravity include the possibility of constructing geodesically complete cosmologies \cite{sc_inv_bars_1}. In this regard see critical comments in Refs. \cite{sc_inv_carrasco, sc_inv_quiros_1, sc_inv_jackiw} and the reply in Ref. \cite{sc_inv_bars_2}. In Refs. \cite{sc_inv_alpha} a new class of chaotic inflation models with spontaneously broken conformal invariance has been developed. In this vein a broad class of multi-field inflationary models with spontaneously broken conformal invariance is described in Ref. \cite{sc_inv_alpha_2}, while generalized versions of these models where the inflaton has a non-minimal coupling to gravity with $\xi<0$, different from its conformal value $\xi=-1/6$, are investigated in Ref. \cite{sc_inv_alpha_3}.

In this paper we avoid semantic issues on what to consider a theory and what a representation of a theory, by following the mainstream of thinking on physical equivalence of different representations of a theory. Hence, physical equivalence is regarded as the inevitable consequence of invariance of the laws of physics -- represented by an action principle and the derived motion equations -- under given transformations; be it coordinate, gauge or conformal transformations. As we shall show, the choice of the above mainstream of thinking has far-reaching consequences not only for the conformal transformations' issue but, also, for the correct understanding of the Weyl-invariant laws of gravity. 

The plan of the paper is as follows. In section \ref{sect-bd-jf-ef} we discuss on the Jordan and Einstein frames of the Brans-Dicke theory. Subsection \ref{subsect-controversy} is dedicated to discuss on the conformal transformations issue on the basis of physical equivalence of the conformal frames, while in subsection \ref{subsect-non-equiv} we briefly discuss on the issue from the point of view of those who assume that the different conformal frames are not physically equivalent. Given that in the present paper we follow the most widespread point of view according to which the necessary requirement for physical equivalence is the invariance of the physical laws under given transformations -- in the present case these are the conformal transformations or Weyl rescalings -- in section \ref{sect-conf-inv-ex} we explore an example of a theory that is indeed conformal invariant and, hence, is well-suited to discuss about physical equivalence of the conformal frames. An aspect of the conformal frames issue that is usually forgotten is related with the geometrical implications of the transformations. In section \ref{sect-sc-inv-geom} we discuss on this issue and we show that Weyl-integrable geometry is better suited than Riemann geometry to discuss on the conformal frames issue. Then, in section \ref{sect-scale-inv} we explore several Weyl-invariant theories of gravity that have been intensively studied in the bibliography, and we make emphasis in the anomalous matter coupling in these theories (only radiation can be consistently coupled), and the related anomaly that is associated with their geometrical basis: While the gravitational laws -- expressed through the action principle and the derived motion equations -- are invariant under the conformal transformations, the geodesics of the metric are indeed transformed into non-geodesic paths. This latter anomaly may be eliminated by considering spacetimes with Weyl-integrable affinity instead of Riemannian spacetimes. This is why in section \ref{sect-consist-weyl} we explore a Weyl-invariant theory of gravity which is associated with Weyl-integrable spacetimes. Further details of this theory, including the coupling of the matter degrees of freedom in a Weyl-invariant way, as well as the consequences of the gauge freedom, are investigated in section \ref{sect-egr}. It is shown, in particular, how to consistently couple the standard model of particles, in such a way as to preserve the Weyl symmetry of the theory. An issue that requires separate discussion is the one about the transformation properties of the constants of the theory. In the bibliography one founds arguments that point to dimensionless constants as being unchanged by the Weyl rescalings, while dimensionful constants should be necessarily transformed. Although at first sign this might appear as a natural statement, one founds dimensionful constants like the Planck constant $\hbar$, the speed of light $c$ and the electric charge of the electron $e$, that are assumed to be unchanged by the Weyl rescalings. This issue is discussed in section \ref{sect-const-nature}, where we conclude that any assumption on the transformation properties of the constants of the theory under the Weyl rescalings, can be either an independent postulate of the theory or a meaningless statement otherwise. Cosmological solutions of the Weyl-invariant theory of gravity are explored in section \ref{sect-cosmo}, while the singularity issue is investigated in section \ref{sect-sing}, in order to illustrate with concrete examples the implications of gauge freedom, as well as of physical equivalence of the different gauges. The results of this paper are discussed in section \ref{sect-discussion} and conclusions are given in section \ref{sec-conclu}. For completeness of our exposition, an appendix section has been included. In the Appendix \ref{sect-conf-t} the transformation of the basic geometric quantities and operators under the conformal transformations is exposed, while in the Appendix \ref{sect-callan} a short exposition on the consequences of quantum effects of matter when the gravitational interactions are considered, is given. In this last section we follow, mainly, the work in Ref. \cite{callan_ann_phys_1970}.

%%%%%%%%%%%%%%%%%%%%%%%%%%%%%%%%%%%%%%%%%%%%%%%%%%%%%%%%%%%%%%%%%%%%%%%%%%%%%%%%%%%%%%%%

\section{Jordan frame and Einstein frame representations of BD theory}\label{sect-bd-jf-ef}

Under the conformal transformation of the metric \eqref{conf-t}, with conformal factor, $\Omega^2=\phi$, together with the rescaling of the BD scalar field: $\phi\rightarrow\exp\vphi$, the Jordan frame BD action:

\bea S^\text{JF}_\text{BD}=\int d^4x\sqrt{|g|}\left[\phi R-\frac{\omega_\text{BD}}{\phi}(\der\phi)^2-2V(\phi)+2{\cal L}_m\right],\label{jf-bd-action}\eea is transformed into the Einstein frame:

\bea &&S^\text{EF}_\text{BD}=\int d^4x\sqrt{|g|}\left[R-\left(\omega_\text{BD}+\frac{3}{2}\right)(\der\vphi)^2\right.\nonumber\\
&&\;\;\;\;\;\;\;\;\;\;\;\;\;\;\;\;\;\;\;\;\;\;\;\;\;\;\;\;\;\;\;\;\;\;\;\;\;\;\;\left.-2V(\vphi)+2e^{-2\vphi}{\cal L}_m\right],\label{ef-bd-action}\eea where, under \eqref{conf-t}: $V(\phi)\rightarrow e^{2\vphi}V(\vphi)$. If consider the boundary term in \eqref{jf-bd-action}, then under the conformal transformation \eqref{conf-t}, it is transformed into its EF counterpart in \eqref{ef-bd-action}: $$2\int_{\der\cal M} d^3x\sqrt{|h|}\phi K\rightarrow 2\int_{\der\cal M} d^3x\sqrt{|h|}K,$$ where $h$ is the determinant of the metric $h_{\mu\nu}$ induced on the boundary and $K=h^{\mu\nu}K_{\mu\nu}$ is its extrinsic curvature scalar. In the latter transformation law for the boundary term it has been taken into account that terms coming from $6\phi\Omega^2\Box(\ln\Omega)$ in the EF action \eqref{ef-bd-action} compensate the terms $-6n^\mu\der_\mu\Omega$ in the EF boundary action (see Appendix A of Ref. \cite{copeland-wands-rev}).

Under \eqref{conf-t} the stress-energy tensor of matter transforms in the following way:

\bea T^{(m)}_{\mu\nu}\rightarrow\Omega^2T^{(m)}_{\mu\nu}\;\;\Rightarrow\,\;T^{(m)}\rightarrow\Omega^4T^{(m)},\label{conf-t-set}\eea while the conservation equation that takes place in the JF transforms into non-conservation equation in the EF:

\bea \nabla^\mu T^{(m)}_{\mu\nu}=0\;\;\rightarrow\;\;\nabla^\mu T^{(m)}_{\mu\nu}=-\frac{\der_\nu\Omega}{\Omega}\,T^{(m)}.\label{conf-t-cons-eq}\eea 

The latter transformation property of the conservation equation is reminiscent of the transformation of the geodesics of the metric under \eqref{conf-t}. Actually, under the conformal transformation of the metric the JF geodesic equation (the same as in GR):

\bea \frac{d^2x^\mu}{ds^2}+\left\{^{\;\mu}_{\sigma\lambda}\right\}\frac{dx^\sigma}{ds}\frac{dx^\lambda}{ds}=0,\label{jf-geod-eq}\eea is transformed into the following EF equation of motion,

\bea &&\frac{d^2x^\mu}{ds^2}+\left\{^{\;\mu}_{\sigma\lambda}\right\}\frac{dx^\sigma}{ds}\frac{dx^\lambda}{ds}=\frac{dx^\sigma}{ds}\frac{dx^\mu}{ds}\der_\sigma\left(\ln\Omega\right)\nonumber\\
&&\;\;\;\;\;\;\;\;\;\;\;\;\;\;\;\;\;\;\;\;\;\;\;\;\;\;\;\;\;\;\;\;\;\;\;\;-g_{\sigma\lambda}\frac{dx^\sigma}{ds}\frac{dx^\lambda}{ds}\der^\mu\left(\ln\Omega\right).\label{ef-geod-eq}\eea It can be shown that under the affine reparametrization: $ds\rightarrow f^{-1}(\Omega)d\tau$, with $f(\Omega)=\Omega$, the first term in the right-hand-side (RHS) of \eqref{ef-geod-eq} can be eliminated, 

\bea \frac{d^2x^\mu}{d\tau^2}+\left\{^{\;\mu}_{\sigma\lambda}\right\}\frac{dx^\sigma}{d\tau}\frac{dx^\lambda}{d\tau}=-g_{\sigma\lambda}\frac{dx^\sigma}{d\tau}\frac{dx^\lambda}{d\tau}\der^\mu\left(\ln\Omega\right).\label{ef-geod-eq'}\eea However, the second term in the RHS of \eqref{ef-geod-eq} can not be eliminated by any affine transformation whatsoever \cite{quiros_grg_2013}. What this means is that in the EFBD there is a universal fifth-force, $$f^\mu_\text{fifth}=-g_{\sigma\lambda}\frac{dx^\sigma}{d\tau}\frac{dx^\lambda}{d\tau}\der^\mu\left(\ln\Omega\right),$$ that deviates the motion of a given particle from being geodesic. A distinctive feature of this universal fifth-force effect is that it acts only on particles with the mass. For massless particles like the photons, gravitons, etc., that move at the speed of light, $g_{\sigma\lambda}dx^\sigma dx^\lambda=0$, so that $f^\mu_\text{fifth}=0$, i. e., massless particles move along geodesics of the metric. The same conclusion is evident from the continuity (non-conservation) equation in the right-hand of \eqref{conf-t-cons-eq}, where it is apparent that for a fluid of massless particles, since $T^{(m)}=0$, then the conservation equation is preserved under the conformal transformation \eqref{conf-t}.

Given that under \eqref{conf-t} the JFBD action \eqref{jf-bd-action} is transformed into the obviously different EFBD action \eqref{ef-bd-action}, while the stress-energy tensor's conservation equation in the JF is transformed into the non-conservation equation in the EF -- see Eq. \eqref{conf-t-cons-eq} -- is is evident that the Brans-Dicke theory, as well as any-other STT, are not invariant under the conformal transformation and, consequently, the invoked invariance of the laws of physics (the gravitational laws being a particular case) is not built-in in these theories. Hence, if one truly believes in the conformal invariance of the physical laws, regarded as invariance under transformations of the physical units in Dicke's sense \cite{dicke-1962}, one is inevitably led to consider other theories of gravity that really embody the conformal symmetry. In spite of this, most part of the bibliography on the conformal transformations' issue avoids a serious discussion of this kind of arguments and the different authors prefer to ``beat around the bush''.

%%%%%%%%%%%%%%%%%%%%%%%%%%%%%%%%%%%%%%%%%%%%%%%%%%%%%%%%%%%%%%%%%%%%%%%%%%%%%%%%%%%%%%%%%%%%%%%%%%%%%%%%%%%%%%%%%%%%%%%%%%%%%%%%%%%%%%%%%%%%%%%%%%%%

\subsection{The conformal transformations' controversy: When the conformal frames are regarded as physically equivalent}\label{subsect-controversy}

Above we have demonstrated a very well known fact: While the JFBD theory \eqref{jf-bd-action} is a STT of gravity in the sense that the gravitational interactions are carried by the metric field of geometric origin, together with the non-geometric BD scalar field, the EFBD theory \eqref{ef-bd-action} is a purely geometric theory of gravity indistinguishable from general relativity but for the presence of an additional non-gravitational universal interaction (fifth-force) between the scalar and the remaining matter fields through the interaction term $\propto e^{-2\vphi}{\cal L}_m$ in the action. This is seen from comparison of the geodesic equations in the Jordan frame \eqref{jf-geod-eq} with the corresponding EF motion equations \eqref{ef-geod-eq}, that can not be reduced to a geodesic by any redefinition whatsoever of the affine parameter. Hence, apparently, JFBD and EFBD are to be regarded as different theories and not as physically equivalent representations of a same theory. 

In spite of the apparent clarity of the above argument, we recommend to read the discussion on this subject in Ref. \cite{sotiriou_etall_ijmpd_2008} (see specially sections 3, 4 and 5 therein) where a different perspective is presented. In that reference one may find interesting arguments that are shared by many researchers. Given that many aspects of the controversy on the physical equivalence of the different conformal representations are reflected in the discussion in that reference, below we cite several selected statements made therein, with the hope that these can help us to better understand the origin of the controversy: 

\begin{enumerate}

\item ``The freedom of having an arbitrary conformal factor is due to the fact that the EEP does not forbid a conformal rescaling in order to arrive at special-relativistic expressions of the physical laws in the local freely falling frame.''

\item ``It should be stressed that all conformal metrics $\phi g_{\mu\nu}$, $\phi$ being the conformal factor, can be used to write down the equations or the action of the theory.''

\item ``As pointed out ... any metric theory can perfectly well be given a representation that appears to violate the metric postulates (recall, for instance, that $g_{\mu\nu}$ is a member of a family of conformal metrics and that there is no {\it a priori} reason why this metric should be used to write down the field equations)''

\item ``... many misconceptions arise when a theory is identified with one of its representations and other representations are implicitly treated as different theories.''

\item ``... the arbitrariness that inevitably exists in choosing the physical variables is bound to affect the representation.''

\item ``Thus, there will be representations in which it will be obvious that a certain principle is satisfied and others in which it will be more intricate to see that. However, it is clear that the theory is one and the same and that the axioms or principles are independent of the representation.''

\item ``This situation is very similar to a gauge theory in which one must be careful to derive only gauge-independent results. Every gauge is an admissible ``representation'' of the theory, but only gauge-invariant quantities should be computed for comparison with experiment. In the case of scalar-tensor gravity, however, it is not clear what a ``gauge'' is and how to identify the analog of ``gauge-independent'' quantities.''

\end{enumerate} 

Which is the missing argument in the above listed statements? The statements in the first three items above, for instance, are all related with the existence of a class of conformal metrics, $\Omega^2g_{\mu\nu}$. But, the BD theory like any other STT is not invariant under the conformal transformations so that this equivalence class is not well-suited to this theory. As a counterexample to this, in the next and the subsequent sections we shall present the action (and the corresponding motion equations) of theories that really embody the conformal invariance, so that the existence of an equivalence class of conformal metrics, $\Omega^2g_{\mu\nu}$, is a natural consequence. A necessary requirement for the existence of this equivalence class is that the scalar field is not determined by a motion equation, i. e., it should be non-dynamical. This is, precisely, the price to pay for conformal invariance, since then, in addition to the four degrees of freedom to make coordinate transformations one has an additional degree of freedom to make conformal transformations of the metric. Contrary to this, in the BD theory -- like in any other STT -- there is always a motion equation that governs the dynamics of the scalar field. This means that, by solving the motion equations one is able to determine not only the spacetime metric, but also the scalar field, so that there is not any freedom in choosing $\phi$. Hence, the missing argument in the reasoning line displayed by the above listed statements is the need for conformal invariance of the equations of the theory (the action plus the derived motion equations) in order to accommodate an equivalence class of conformal metrics. 

We shall not discuss on the arguments displayed in the items 4, 5 and 6 above, since these are highly dependent on the assumed definition of what to understand by a representation of a theory and, hence, we shall inevitably end up discussing on semantic issues that have nothing to do with the core of the conformal frames' controversy. However, the statement in the last (seventh) item needs of some discussion. This statement contradicts the viewpoint on physical equivalence of the conformal frames advocated in Ref. \cite{sotiriou_etall_ijmpd_2008}. According to the authors, the situation on physical equivalence of the conformal frames should be compared with a gauge theory -- as we have discussed before in the example given by the Lagrangian \eqref{gauge-lag} -- where, although every gauge is an admissible representation of the theory, only gauge-invariant quantities are of relevance for purposes of comparison with the experiment. The authors themselves recognize in a sentence of this last statement that, in the case of scalar-tensor theories ``it is not clear what a gauge is and how to identify the analog of gauge-independent quantities''.\footnote{In Ref. \cite{jarv} a formalism was developed that allows to construct the invariants that are to be linked with measurable quantities in the STT. In this regard we want to make a comment: Attaching physical (measurable) meaning to gauge invariant quantities in a theory that is not itself gauge invariant, makes sense only as an additional postulate, so that one ends up dealing with a completely different theory (not a STT in the standard sense).} In view of our adopted concept of physical equivalence of the conformal frames as related with conformal symmetry of the gravitational laws (and the remaining laws of physics), the lack of clarity in what to understand by a gauge and what to identify by gauge-independent quantities within the framework of the STT-s, in connection with conformal transformations, is due to the fact that these theories do not actually embody conformal invariance.

%%%%%%%%%%%%%%%%%%%%%%%%%%%%%%%%%%%%%%%%%%%%%%%%%%%%%%%%%%%%%%%%%%%%%%%%%%%%%%%%%%%%%%%%%%%%%%%%%%%%%%%%%%%%%%

\subsection{When the JF and EF representations are regarded as non-equivalent frameworks}\label{subsect-non-equiv}

Up to this point in our discussion we have considered the controversy that arises when the different conformal frames are considered as physically equivalent representations of a same theory. In this subsection we shall briefly expose the point of view according to which the different conformal frames are regarded as physically non-equivalent representations, i. e., here we shall be concerned with the second of the questions that stand at the core of the conformal frames' controversy stated at the beginning of the introductory section. For instance, in the Refs. \cite{faraoni_ijtp_1999, sarkar_mpla_2007, ct-ineq-nojiri, ct-ineq-brisc, ct-ineq-capoz, ct-ineq-brooker, ct-ineq-baha, ct(inequiv)-7, ct(inequiv)-8, ct(inequiv)-9} the physical equivalence of the JF and EF conformal frames is challenged both classically \cite{faraoni_ijtp_1999, sarkar_mpla_2007, ct-ineq-nojiri, ct-ineq-brisc, ct-ineq-capoz, ct-ineq-brooker, ct-ineq-baha} and at the quantum level \cite{ct(inequiv)-7, ct(inequiv)-8, ct(inequiv)-9}. In Ref. \cite{faraoni_ijtp_1999} an example based on gravitational waves is explored in order to clarify the issue. It is seemingly demonstrated therein that the EF is the better suited frame to describe the physical phenomena. It has been shown in Ref. \cite{sarkar_mpla_2007} that the gravitational deflection of light to second order accuracy may observationally distinguish the two conformally related frames of the BD theory. Meanwhile in Refs. \cite{ct-ineq-nojiri, ct-ineq-brisc, ct-ineq-capoz, ct-ineq-brooker, ct-ineq-baha}, by means of the equivalence between the $f(R)$ and STT theories, the physical non-equivalence of the JF and EF frames is demonstrated. The non-equivalence of these formulations of the BD theory from the physical standpoint has been investigated also in Refs. \cite{kaloper_prd_1998, quiros_prd_2000} in what regards to the spacetime singularities. 

Here, as before, if appropriate care is not taken about involved concepts, the discussion may go on to a semantic issue. First, what means that the different conformal frames are physically non-equivalent? After all, when one compares two different frames, even when these are related by a mathematical relationship of equivalence, as long as the physical laws are not invariant under the equivalence relationship, what one is comparing is two different theories with their own set of laws and of measurable quantities. Hence, it is natural to get different predictions for a given quantity when computed in terms of the measurable quantities of one or another frame. In this regard, looking for evidence on the non-equivalence of the different conformal frames amounts to looking for evidence in favor of one or the other theoretical framework, no more. This is precisely the point we want to make here in what regards the different conformal frames of BD theory and of STT.

A different thing is to search for a physical conformal frame among the conformally related ones. This would be a task inevitably doomed to failure. Actually, if the conformally related frames are physically equivalent, then all (or none) of them are physical. If they are not physically equivalent, then the different frames represent actually different theories: for instance JFBD is a metric STT while EFBD is GR supplemented with an additional non-gravitational universal fifth-force, i. e., a non-metric theory. In this case what matters is not whether the theory is physical or not but whether the theory's predictions meet or not the experimental evidence. Nevertheless one founds statements like this (here we do not cite any particular work since this kind of statement is quite generalized among researchers): ``... the matter is coupled to the conformal metric $\Omega^2g_{\mu\nu}$ (the physical metric) and not to the gravitational metric $g_{\mu\nu}$.'' It is not difficult to understand that, in such cases when one may differentiate the gravitational metric from a metric to which the matter is coupled -- which in such kind of statement means that the latter is the metric in terms of which the stress-energy tensor of matter is conserved -- what one has is not a STT, nor even GR, but a bimetric theory of gravity. That this is not usually recognized is just an indication of the lack of understanding of what a conformal transformation of the metric really entails for the STT-s.

%%%%%%%%%%%%%%%%%%%%%%%%%%%%%%%%%%%%%%%%%%%%%%%%%%%%%%%%%%%%%%%%%%%%%%%%%%%%%%%%

\section{An example of a conformal invariant BD theory}\label{sect-conf-inv-ex}

If one follows -- as we do here -- the mainstream of thinking regarding physical equivalence of different representations of a theory, according to which the physical equivalence of the different representations is necessarily linked with invariance of the laws of physics -- represented by an action principle and the derived motion equations -- under given transformations; be it coordinate, gauge or conformal transformations, one is inevitably led to consider theories that really embody the conformal symmetry.

An example of a STT theory which is invariant under the conformal transformation of the metric plus a redefinition of the scalar field, known as Weyl rescalings: 

\bea g_{\mu\nu}\rightarrow\Omega^{-2}g_{\mu\nu},\;\;\vphi\rightarrow\vphi+2\ln\Omega,\label{weyl-t}\eea is given by the following vacuum action \cite{quiros_grg_2013}: 

\bea S=\int d^4x\sqrt{|g|} e^\vphi\left[R+\frac{3}{2}(\der\vphi)^2\right],\label{conf-inv-action}\eea and the corresponding field equations,

\bea &&G_{\mu\nu}=-\frac{1}{2}\left[\der_\mu\vphi\der_\nu\vphi+\frac{1}{2}g_{\mu\nu}(\der\vphi)^2\right]\nonumber\\
&&\;\;\;\;\;\;\;\;\;\;\;\;\;\;\;\;\;\;\;\;\;\;+\left(\nabla_\mu\nabla_\nu-g_{\mu\nu}\Box\right)\vphi,\nonumber\\
&&\Box\vphi+\frac{1}{2}(\der\vphi)^2-\frac{1}{3}R=0.\label{conf-inv-feqs}\eea The above equations (including the action itself) are not transformed by \eqref{weyl-t}. Besides, the motion equation for the scalar field is not independent from the Einstein's equation, since the trace of the first equation in \eqref{conf-inv-feqs} coincides with the second (KG) equation. This means that there is not an independent equation that governs the dynamics of $\vphi$, so that, the scalar field in this theory -- as in any other conformal invariant theory -- is not a dynamical degree of freedom. This is an inevitable consequence of conformal invariance. Actually, besides the four degrees of freedom to make diffeomorphisms, there is one more degree of freedom to make conformal transformations, so that we can set the scalar field $\vphi$ to any function we want. As an example of this freedom, let us to look for cosmological solutions of this theory under the assumption of Friedmann-Robertson-Walker (FRW) spacetime background with flat spatial sections, whose line-element is given by: $ds^2=-dt^2+a^2(t)\delta_{ik}dx^idx^k,$ ($i,k=1,2,3$). The corresponding Friedmann and Raychaudhuri equations in \eqref{conf-inv-feqs} read (the KG equation is not independent from these):

\bea 3H^2&=&-\frac{3}{4}\dot\vphi^2-3H\dot\vphi,\nonumber\\
-2\dot H&=&-\frac{1}{2}\dot\vphi^2-H\dot\vphi+\ddot\vphi.\nonumber\eea The first equation above can be written as: $3(H+\dot\vphi/2)^2=0$, while the second one amounts to: $\dot H+\ddot\vphi/2=0$, which is not independent from the former one. Hence, as solution of the equation of motion, one is left with a relationship between the scale factor and the gauge field: $a(\vphi)=a_0\exp(-\vphi/2),$ where $a_0=\exp C_0$ ($C_0$ is an integration constant). One is then free to choose any function $\vphi=\vphi(t)$ one wants, that would in turn drive a desired cosmic dynamics through $a=a(\vphi(t))$. A similar analysis can be found in Ref. \cite{alvarez-jcap-2015} (see, in particular, section 3 of this reference) and in \cite{sc_inv_quiros} (see section VI). 

The invariance of the motion equations \eqref{conf-inv-feqs} -- and of the action \eqref{conf-inv-action} -- of the theory under the Weyl rescalings \eqref{weyl-t} entails that, instead of a given metric tensor $g_{\mu\nu}$, one has a whole class of conformal metrics, $\Omega^2g_{\mu\nu}$, through which one can geometrically interpret the physical (gravitational) phenomena. In other words, we have at our disposal a class of equivalent geometrical ``realizations'' of given physical laws. These different geometrical realizations is what we may consider as physically equivalent representations of the theory. 

In order to better illustrate the analysis, let us assume that the pair, $(g^{(0)}_{\mu\nu},\vphi_{(0)})$, where $g^{(0)}_{\mu\nu}=g^{(0)}_{\mu\nu}(x)$ and $\vphi_{(0)}=\vphi_{(0)}(x)$ are point-dependent functions, accounts for any given -- ``starting'' -- representation of the theory \eqref{conf-inv-action}. By means of the conformal transformation \eqref{weyl-t} with a chosen specific function $\Omega^2_{(k)}=f_{(k)}(\vphi_{(0)})$ ($f_{(k)}$ is a positive continuous function), from this starting representation of the theory, a new representation is obtained: $$g^{(k)}_{\mu\nu}=\Omega^2_{(k)}g^{(0)}_{\mu\nu},\;\;\vphi_{(k)}=\vphi_{(0)}-2\ln\Omega_{(k)}.$$ Since both pairs, $(g^{(0)}_{\mu\nu},\vphi_{(0)})$, and $(g^{(k)}_{\mu\nu},\vphi_{(k)})$, obey the same equations of motion \eqref{conf-inv-feqs} (these obey also, of course, the same action principle \eqref{conf-inv-action}), we can say that these pairs amount to two different but physically equivalent representations of the same theory. Here one may define conformal invariant (also coordinate invariant) quantities that can be related to the measurables of the theory as, for instance, $d\tau^2_\text{inv}:=e^\vphi d\tau^2$ ($d\tau$ is the coordinate invariant time interval), $R_\text{inv}:=e^{-\vphi}[R+3(\der\vphi)^2/2]$, where $R$ is the curvature scalar, etc.

The problem with the theory \eqref{conf-inv-action} is that, excluding traceless matter, the remaining matter degrees of freedom can not be consistently coupled to the theory since the trace of the Einstein's equation for vacuum coincides with the Klein-Gordon equation. This fact would be immediately understood by the reader if we would mention from the start that the action above is nothing but Brans-Dicke theory with the special value of the coupling constant \cite{deser}: $\omega_\text{BD}=-3/2$. Regardless of this, it serves as an example of a theory that really embodies the conformal invariance of the laws of physics invoked in Refs. \cite{faraoni_prd_2007, dicke-1962} and related work.

%%%%%%%%%%%%%%%%%%%%%%%%%%%%%%%%%%%%%%%%%%%%%%%%%%%%%%%%%%%%%%%%%%%%%%%%%%%%%%%%%%%%%%%%%%%%

\section{The (forgotten) geometrical aspect of conformal invariance}\label{sect-sc-inv-geom}

A less known aspect of the conformal transformations of STT, in particular of BD theory, has been explored in Ref. \cite{quiros_grg_2013} (see also Refs. \cite{quiros_prd_2000, quiros_npb_2002, scholz, sc_inv_scholz_1, sc_inv_scholz_2, romero, pucheu, lobo}). It is linked with the geometrical face of the conformal transformations \eqref{conf-t}. According to the line of reasoning in Ref. \cite{quiros_grg_2013}, under the conformal transformation \eqref{conf-t}, the transformation of the Christoffel symbols of the metric $g_{\mu\nu}$ in \eqref{aff-conf-t} may be interpreted in the following alternative way:

\bea \{^\sigma_{\mu\nu}\}\rightarrow\Gamma^\sigma_{\mu\nu}\equiv\{^\sigma_{\mu\nu}\}+\frac{1}{2}\left(\delta^\sigma_\mu\der_\nu\vphi+\delta^\sigma_\nu\der_\mu\vphi-g_{\mu\nu}\der^\sigma\vphi\right),\label{alt-way}\eea where $\Gamma^\sigma_{\mu\nu}$ are the affine connection of Weyl-integrable geometry\footnote{This is a particular case in the class of a the more general Weyl geometries \cite{weyl}.} (WIG) with the Weyl gauge scalar being identified with the logarithm of the conformal factor: $\vphi\equiv\ln\Omega^2$. The metricity condition of the WIG (the supra-index $(w)$ means that given quantities and operators are defined in terms of the affine connection $\Gamma^\sigma_{\mu\nu}$) is given by: $\nabla^{(w)}_\sigma g_{\mu\nu}=-\der_\sigma\vphi g_{\mu\nu}.$ It has been proposed in \cite{quiros_grg_2013} that the geometrical structure better suited to address conformal invariance or invariance under Weyl rescalings is not (pseudo)Riemann geometry -- as it is implicitly assumed when the issue is discussed in the bibliography -- but WIG instead (see section \ref{sect-consist-weyl} below). 

If assume WIG as the geometrical structure to be associated with the action \eqref{conf-inv-action}, the conformal invariance of the latter theory is complemented with the conformal invariance of the associated geometrical background, including invariance of the WIG geodesics under \eqref{weyl-t}. As a consequence, the issue of the conformal frames acquires a new dimension if assume the alternative way \eqref{alt-way}. Actually, in this case a conformal transformation from the JF to the EF takes us from (pseudo)Riemannian manifold into a WIG-space. Hence, from the start it is not required to compare these frames since these are associated with different geometrical structures. For a more detailed discussion on this new aspect of the conformal transformation's issue we recommend Ref. \cite{quiros_grg_2013}.

%%%%%%%%%%%%%%%%%%%%%%%%%%%%%%%%%%%%%%%%%%%%%%%%%%%%%%%%%%%%%%%%%%%%%%%%%%%%

\section{Anomalous Weyl-invariant theories of gravity}\label{sect-scale-inv}

Invariance of the laws of gravity under Weyl rescalings \eqref{weyl-t} has been repeatedly invoked in the context of the SMP coupled to gravity \cite{smolin_npb_1979, zee_prd_1981, cheng_prl_1988, goldberger_prl_2008, sc_inv_indios, sc_inv_indios_1, sc_inv_shapo, sc_inv_percacci, sc_inv_prester, sc_inv_quiros, sc_inv_padilla, sc_inv_bars, sc_inv_bars_1, sc_inv_bars_2, sc_inv_carrasco, sc_inv_quiros_1, sc_inv_jackiw, sc_inv_alpha, sc_inv_alpha_2, sc_inv_alpha_3, sc_inv_alpha_4, salvio-1, sc_inv_ghorbani, sc_inv_farz, sc_inv_vanzo, sc_inv_alvarez, alvarez-jcap-2015, sc_inv_khoze, sc_inv_karananas, sc_inv_tambalo, sc_inv_kannike, sc_inv_ferreira, sc_inv_maeder, sc_inv_myung}. In order to discuss on local scale invariance of the gravitational laws it is useful to write the following prototype action \cite{deser, smolin_npb_1979}:

\bea S=\int d^4x\sqrt{|g|}\left[\frac{\phi^2}{12}\,R+\frac{1}{2}(\der\phi)^2\pm\frac{\lambda}{12}\,\phi^4\right].\label{deser-action}\eea Since, under the Weyl rescalings:

\bea g_{\mu\nu}\rightarrow\Omega^{-2}g_{\mu\nu},\;\phi\rightarrow\Omega\,\phi,\label{scale-t}\eea the combination $\sqrt{|g|}[\phi^2R+6(\der\phi)^2]$ is kept unchanged -- as well as the scalar density $\sqrt{|g|}\phi^4$ -- then the action (\ref{deser-action}) is invariant under \eqref{scale-t}. Any scalar field which appears in the gravitational action the way $\phi$ does, is said to be conformally coupled to gravity. Hence, for instance, the following action \cite{sc_inv_prester, sc_inv_bars, sc_inv_bars_1, sc_inv_bars_2, sc_inv_carrasco, sc_inv_quiros_1, sc_inv_jackiw, sc_inv_alpha, sc_inv_alpha_2, sc_inv_alpha_3, sc_inv_alpha_4}:

\bea S=\int d^4x\sqrt{|g|}\left[\frac{\left(\phi^2-\sigma^2\right)}{12}\,R+\frac{1}{2}(\der\phi)^2-\frac{1}{2}(\der\sigma)^2\right],\label{bars-action}\eea is also invariant under \eqref{scale-t} since both $\phi$ and $\sigma$ are conformally coupled to gravity, provided that the additional scalar field $\sigma$ transforms in the same way as $\phi$: $\sigma\rightarrow\Omega\,\sigma$. For the coupling $\propto (\phi^2-\sigma^2)^{-1}$ to be positive and the theory Weyl-invariant, the scalar $\vphi$ must have a wrong sign kinetic energy -- just like in \eqref{deser-action} -- potentially making it a ghost. However, the local Weyl gauge symmetry compensates, thus ensuring the theory is unitary \cite{sc_inv_bars, sc_inv_bars_1, sc_inv_bars_2}.

In the model of Refs. \cite{sc_inv_bars_1, sc_inv_bars_2}, in order to get geodesically complete spacetimes it is required that not only the field $\vphi$, but also the doublet Higgs field $H$ be a set of conformally coupled scalars consistent with $SU(2)\times U(1)$. The corresponding Weyl-invariant action that describes the coupling of gravity and the standard model reads: $S=\int d^4x\sqrt{|g|}\left({\cal L}+{\cal L}_\text{SMP}\right)$, where ${\cal L}_\text{SMP}$ is the Lagrangian of the standard model of particles and,

\bea &&{\cal L}=\frac{1}{12}\left(\phi^2-2H^\dag H\right)R+\frac{1}{2}(\der\phi)^2-|DH|^2\nonumber\\
&&\;\;\;\;\;\;\;\;\;\;\;\;\;\;\;\;\;\;\;\;\;\;-\frac{\lambda}{4}\left(H^\dag H-\alpha^2\phi^2\right)^2-\frac{\lambda'}{4}\phi^4.\label{higgs-bars-action}\eea In the above equation we have used the notation $|DH|^2\equiv g^{\mu\nu}D_\mu H^\dag D_\nu H$, where $D_\mu$ stands for the gauge-covariant derivative. The above action is invariant under the Weyl rescalings \eqref{scale-t} plus the following rescaling of the remaining fields in ${\cal L}$ and in ${\cal L}_\text{SMP}$: $$H\rightarrow\Omega H,\;\;\psi\rightarrow\Omega^{3/2}\psi,\;\;A^a_\mu\rightarrow A^a_\mu,$$ where $\psi$ stands for the fermion fields for quarks or leptons, and the $A^a_\mu$ are the gauge fields for the photon, gluons, $W^\pm$ and $Z^0$ bosons. In this theory the only scale is generated by gauge fixing $\phi$ to a constant: $\phi(x)\rightarrow\phi_0$. All dimensionful parameters emerge from this single source: $$\frac{1}{16\pi G_N}=\frac{\phi_0^2}{12},\;\;\frac{\Lambda}{16\pi G_N}=\lambda'\phi_0^4,\;\;H^\dag_0 H_0=\alpha^2\phi_0^2.$$

In Refs. \cite{sc_inv_alpha, sc_inv_alpha_2, sc_inv_alpha_3, sc_inv_alpha_4}, a new class of conformally invariant theories which allow inflation, even if the scalar potential is very steep in terms of the original conformal variables, was explored. In order to understand how the cosmological attractor arises in these theories let us to investigate a toy model \cite{sc_inv_alpha_4} given by the action \eqref{bars-action} supplemented with the potential term $$-\int d^4x\sqrt{|g|}\frac{\lambda}{36}\left(\phi^2-\sigma^2\right)^2.$$ In addition to the invariance under Weyl rescalings, $$g_{\mu\nu}\rightarrow\Omega^{-2}g_{\mu\nu},\;\;(\phi,\sigma)\rightarrow\Omega(\phi,\sigma),$$ it has a global $SO(1,1)$ symmetry with respect to a boost between the two fields $\phi$, $\sigma$, preserving the value of $\phi^2-\sigma^2$, which resembles the Lorentz symmetry of the theory of special relativity. Since, $\phi^2-\sigma^2>0$ in order to have gravity rather than antigravity, $\phi$ represents a cutoff for the possible values of $\sigma$. Notice, however, that $\phi$ is not a physical degree of freedom since it may be gauged away, for instance, by fixing the gauge: $\phi^2-\sigma^2=6$, so that $$\phi=\sqrt{6}\cosh(\vphi/\sqrt{6}),\,\;\sigma=\sqrt{6}\sinh(\vphi/\sqrt{6}).$$ Under this choice of the gauge, the starting action is transformed into the following GR action with a canonical (minimally coupled) scalar field and a cosmological constant: $$S=\int d^4x\sqrt{|g|}\left[\frac{1}{2}R-\frac{1}{2}(\der\vphi)^2-\lambda\right].$$ The potential term $\propto(\phi^2-\sigma^2)^2$ is the ``placeholder'' for what becomes a cosmological constant. The main idea developed in Refs. \cite{sc_inv_alpha, sc_inv_alpha_2, sc_inv_alpha_3, sc_inv_alpha_4} is that one can construct a class of inflationary models by locally modifying the would be cosmological constant (the placeholder): 

\bea &&-\int d^4x\sqrt{|g|}\frac{\lambda}{36}\left(\phi^2-\sigma^2\right)^2\nonumber\\
&&\;\;\;\;\;\;\;\;\;\;\;\;\;\;\;\;\rightarrow-\int d^4x\sqrt{|g|}\frac{1}{36}f^2(\sigma/\phi)\left(\phi^2-\sigma^2\right)^2,\nonumber\eea where $f^2(\sigma/\phi)$ is an arbitrary function of the ratio $\sigma/\phi$. Through $f^2(\sigma/\phi)$ one deforms the starting $SO(1,1)$ symmetry. In the gauge $\phi^2-\sigma^2=6$, one gets: $$S=\int d^4x\sqrt{|g|}\left[\frac{1}{2}R-\frac{1}{2}(\der\vphi)^2-f^2(\tanh\vphi/\sqrt{6})\right].$$ Hence, since asymptotically $\tanh\vphi/\sqrt{6}\rightarrow\pm 1$, i. e., $f^2(\tanh\vphi/\sqrt{6})\rightarrow$ const, the system in the large $\vphi$-limit evolves asymptotically towards its critical point where the $SO(1,1)$ symmetry is restored.

%-----------------------------------------------------------------

\subsection{Anomalous matter coupling}\label{subsect-anomalous-c}

The motion equations that can be derived from \eqref{deser-action} (here for simplicity we drop the potential term $\propto\phi^4$) read:

\bea &&\phi^2G_{\mu\nu}=-4\der_\mu\phi\der_\nu\phi+g_{\mu\nu}(\der\phi)^2\nonumber\\
&&\;\;\;\;\;\;\;\;\;\;\;\;\;\;+2\phi\left(\nabla_\mu\nabla_\nu-g_{\mu\nu}\Box\right)\phi,\nonumber\\
&&\Box\phi-\frac{1}{6}\,R\phi=0.\label{deser-feqs}\eea The interesting thing is that the trace of the first equation above (the Einstein's equation) exactly coincides with the KG motion equation that is derived from \eqref{deser-action}. Hence, if add minimally coupled matter with stress-energy tensor $T_{\mu\nu}^{(m)}$ (it would appear with a factor of $6$ in the RHS of the first equation), the trace of the Einstein's equation then would yield: $$\Box\phi-\frac{1}{6}\,R\phi=\frac{1}{\phi}T^{(m)},$$ while the KG equation continues being given by the second equation in \eqref{deser-feqs}. Hence, only traceless matter: $T^{(m)}=0$, can be consistently coupled in the above scale invariant theory of gravity, unless one allows the matter to couple non-minimally \cite{sc_inv_quiros}. This result is easily understood if note that under the scalar field replacement, $\phi\rightarrow\phi^2$, up to the factor of $1/12$, the action \eqref{deser-action} is just BD theory with the special (anomalous) value of the coupling constant ($\omega_\text{BD}=-3/2$) \cite{deser}. This problem with the anomalous coupling of matter in the theory \eqref{deser-action}, as well as in \eqref{bars-action}, is usually misunderstood or just evaded. 

In addition to the above problem, as shown in Ref. \cite{sc_inv_jackiw}, Weyl invariance of the theory \eqref{deser-action} -- the same for \eqref{bars-action} -- does not have any dynamical role since its associated Noether symmetry current vanishes. Besides, another (perhaps related) aspect of local scale invariance of gravity theories, that is not discussed as frequently as desired, is related with its geometrical implications: Conformal invariance or invariance under Weyl rescalings \eqref{scale-t}, is meaningless until a geometrical background is specified. Here by geometrical background we do not understand just a metric but a whole geometrical set up. I. e., a set of geometrical laws that define a geometrical structure, for instance, (pseudo)Riemann geometry, or Weyl geometry, etc. Usually it is implicitly assumed that the background geometry is (pseudo)Riemann, but, in what regards conformal invariance, this implicit choice has its own drawbacks. As demonstrated in section \ref{sect-bd-jf-ef}, under a conformal transformation of the metric in \eqref{scale-t} the geodesics of the metric are transformed into non-geodesics in the conformal frame. Hence, assuming the action \eqref{deser-action} to be defined on a pseudo-Riemann manifold, means that, while the gravitational laws -- represented by the action and the derived equations of motion -- are indeed invariant under the Weyl rescalings \eqref{scale-t}, the geodesics of the metric are transformed into non-geodesics paths. This means, in turn, that there exists an anomalous fifth-force effect in one of the conformally related representations given that it is absent in the other one. This effect, by itself, invalidates the assumed Weyl invariance of the laws of gravity in the theory \eqref{deser-action} and/or in \eqref{bars-action}, since the ``gauge'' field $\phi$ becomes into a dynamical degree of freedom, that is incompatible with local scale invariance. We think that this is, precisely, the origin of the non-dynamical property of the Weyl invariance discovered in \cite{sc_inv_jackiw}.

%%%%%%%%%%%%%%%%%%%%%%%%%%%%%%%%%%%%%%%%%%%%%%%%%%%%%%%%%%%%%%%%%%%%%%%%%%%%%%%%%

\section{Anomaly-free Weyl-invariant theory of gravity}\label{sect-consist-weyl}

As discussed in section \ref{sect-sc-inv-geom}, a way out of the above discussed problem may be to look for an appropriate geometrical structure that may really embody the conformal invariance of the action -- and of the derived motion equations -- of the theory. Here we propose that this geometric structure is a Weyl-integrable manifold (see also \cite{quiros_grg_2013, sc_inv_quiros, sc_inv_novello, sc_inv_scholz_1, quiros_prd_2000, scholz, romero, lobo, pucheu}). 

Let us assume that the background geometry is Weyl-integrable\footnote{For a compact introduction to Weyl geometry in general and to Weyl-integrable geometry as its ``healthy'' particular case, see section 2 of Ref. \cite{quiros_grg_2013}. For a phylosophical and historical perspective on this topic we recommend \cite{sc_inv_scholz_1, sc_inv_scholz_2}.} instead of pseudo-Riemann. In this case the affine connection of the geometry does not coincide with the Christoffel symbols of the metric but it is given, instead, by:

\bea \Gamma^\sigma_{\mu\nu}\equiv\{^\sigma_{\mu\nu}\}+\frac{1}{2}\left(\delta^\sigma_\mu\der_\nu\vphi+\delta^\sigma_\nu\der_\mu\vphi-g_{\mu\nu}\der^\sigma\vphi\right),\label{weyl-aff-c}\eea where $\vphi$ is the Weyl gauge boson which here is not related to the conformal factor as in \eqref{alt-way}. Under the Weyl rescalings \eqref{weyl-t}: $$g_{\mu\nu}\rightarrow\Omega^{-2}g_{\mu\nu},\;\;\vphi\rightarrow\vphi+2\ln\Omega,$$ the above affine connection is not transformed; $\Gamma^\sigma_{\mu\nu}\rightarrow\Gamma^\sigma_{\mu\nu}$, while: $$R^{(w)}_{\mu\nu}\rightarrow R^{(w)}_{\mu\nu},\;\;R^{(w)}\rightarrow\Omega^2R^{(w)}\Rightarrow G^{(w)}_{\mu\nu}\rightarrow G^{(w)}_{\mu\nu},$$ where the quantities (and operators) with supra-index $(w)$ are defined in terms of the affine connection \eqref{weyl-aff-c} so that, for instance, $G^{(w)}_{\mu\nu}$ stands for the Weyl-integrable Einstein's tensor, etc. Hence, thinking along these lines, one may replace the BD action \eqref{conf-inv-action} with the anomalous value of the coupling constant ($\omega_\text{BD}=-3/2$), by its Weyl-integrable (also Weyl-invariant) counterpart \cite{sc_inv_quiros}: 

\bea S_\text{WI}=\frac{1}{2}\int d^4x\sqrt{|g|}e^\vphi R^{(w)}.\label{weyl-inv-action}\eea 

In this theory the scalar field $\vphi$ is a gauge field, i. e., it is not a dynamical degree of freedom since it can be safely gauged away without any physical consequences. If in the RHS of \eqref{weyl-inv-action} add a matter piece of action $S_m=\int d^4x\sqrt{|g|}{\cal L}_m$, the derived Weyl-integrable Einstein's equation \cite{sc_inv_quiros}: 

\bea G^{(w)}_{\mu\nu}=e^{-\vphi}T^{(m)}_{\mu\nu},\label{wim-feq}\eea is not only Weyl-invariant, but also admits coupling of matter degrees of freedom other than the radiation (see the next section). Here we want to notice that in the matter action, the Lagrangian density $\sqrt{|g|}{\cal L}_m$ is not transformed under the Weyl rescalings \eqref{weyl-t} since, under the conformal transformation: $g_{\mu\nu}\rightarrow\Omega^{-2}g_{\mu\nu}$, ${\cal L}_m\rightarrow\Omega^4{\cal L}_m$, while $\sqrt{|g|}\rightarrow\Omega^{-4}\sqrt{|g|}$.

One of the first Weyl-invariant theories of gravity with the SMP minimally coupled was proposed in Ref. \cite{smolin_npb_1979} and then a related model was proposed in Ref. \cite{cheng_prl_1988}. In these proposals the Weyl geometry was assumed so that, instead of a Weyl-gauge scalar as above, a Weyl-gauge vector played the role of the gravitational gauge field. In the theory of \cite{cheng_prl_1988} the EW symmetry breaking potential not only allows for generation of masses of the gauge bosons (and fermions) but, also, generates the Planck mass. The requirement of local scale invariance of this theory leads to the existence of Weyl's vector meson which absorbs the Higgs particle remaining in the Weinberg-Salam model. Here, for simplicity and in order to avoid certain issues associated with the assumption of a gauge vector, we have decided to choose the simplest Weyl-integrable geometrical structure with its associated gauge scalar $\vphi$ instead.

%%%%%%%%%%%%%%%%%%%%%%%%%%%%%%%%%%%%%%%%%%%%%%%%%%%%%%

\section{Extended general relativity}\label{sect-egr}

In what follows we shall explore the physical consequences of the simplest of the Weyl-invariant theories of gravity, that is given by the following action and the corresponding field equations:\footnote{More complex Weyl-invariant Lagrangians may include higher-order curvature terms like: $${\cal L}_\text{high}=\left[aR^2_{(w)}+b R^{(w)}_{\mu\nu}R_{(w)}^{\mu\nu}+c e^{-\vphi}R^{(w)}_{\mu\nu\tau\lambda}R_{(w)}^{\mu\nu\tau\lambda}\right],$$ where $a$, $b$ and $c$ are constants. The following transformation properties of given quantities under \eqref{weyl-t}, have been considered: $$R^2_{(w)}\rightarrow\Omega^4R^2_{(w)},\;R^{(w)}_{\mu\nu}\rightarrow R^{(w)}_{\mu\nu},\;R_{(w)}^{\mu\nu}\rightarrow\Omega^4R_{(w)}^{\mu\nu},$$ $$R^{(w)}_{\mu\nu\tau\lambda}\rightarrow\Omega^{-2}R^{(w)}_{\mu\nu\tau\lambda},\;R_{(w)}^{\mu\nu\tau\lambda}\rightarrow\Omega^8R_{(w)}^{\mu\nu\tau\lambda}.$$}

\bea &&S=\int d^4x\sqrt{|g|}\left[\frac{M^2_\text{Pl}e^\vphi}{2} R^{(w)}-\lambda e^{2\vphi}+{\cal L}_m\right],\nonumber\\
&&G^{(w)}_{\mu\nu}=\frac{e^{-\vphi}}{M^2_\text{Pl}}\,T^{(m)}_{\mu\nu}-\frac{\lambda e^\vphi}{M^2_\text{Pl}}g_{\mu\nu},\label{weyl-inv-theor}\eea where $\lambda$ is a free constant. How the Planck mass squared -- being a constant with the dimensions of mass squared -- affects the Weyl invariance of the theory, is a subject that will be discussed below in section \ref{sect-const-nature}. For the moment we assume, without discussion, that only the fields are transformed by the Weyl rescalings \eqref{weyl-t}. Hence, the constants of nature, no matter whether these are dimensionless or dimensionful, are not transformed by the Weyl rescalings. 

In the above action the explicitly assumed geometrical structure of the underlying spacetime is WIG. Since in this case not only the orientation, but also the length of vectors, $l=\sqrt{g_{\mu\nu}l^\mu l^\nu}$, are point-dependent quantities: $$\frac{dl}{l}=\frac{1}{2}dx^\mu\der_\mu\vphi\;\;\Leftrightarrow\;\;l=l_0 e^{\vphi/2},$$ this entails that the units of measure themselves are point dependent measures as well. Given that the contracted Bianchi identity in the WIG reads: $\nabla^\mu_{(w)}G^{(w)}_{\mu\nu}=0$, and that the Weyl-integrable metricity condition is given by:

\bea \nabla^\sigma_{(w)}g_{\mu\nu}=-\der^\sigma\vphi g_{\mu\nu},\label{wim-met-cond}\eea then the conservation equation that follows from the Einstein's equation \eqref{weyl-inv-theor} can be written as:

\bea \nabla^\mu_{(w)}T^{(m)}_{\mu\nu}=\der^\mu\vphi\,T^{(m)}_{\mu\nu},\label{conserv-eq}\eea where the RHS is not to be considered as a source term, since it is originated from the fact that the units of measure of the stresses are point-dependent quantities in WIG (compare with the Weyl-integrable metricity condition \eqref{wim-met-cond}). It may be convenient to introduce the Weyl-integrable stress-energy tensor (WISET) for the matter: 

\bea T^{(w,m)}_{\mu\nu}:=e^{-\vphi}T^{(m)}_{\mu\nu},\label{weyl-set}\eea so that the conservation equation \eqref{conserv-eq} takes the ``formal'' conservation equation look: $$\nabla^\mu_{(w)}T^{(w,m)}_{\mu\nu}=0.$$ Notice that the WISET shares the same transformation properties of the Weyl-integrable Einstein's tensor $G^{(w)}_{\mu\nu}$ under the Weyl rescalings \eqref{weyl-t}. In particular, $T^{(w,m)}_{\mu\nu}$, is not transformed by \eqref{weyl-t}.

The Weyl-invariant property of the above conservation equation is reminiscent of the invariance of the WIG geodesics: 

\bea \frac{d^2x^\mu}{ds^2}+\Gamma^\mu_{\sigma\lambda}\frac{dx^\sigma}{ds}\frac{dx^\lambda}{ds}-\frac{1}{2}\der_\lambda\vphi\frac{dx^\lambda}{ds}\frac{dx^\mu}{ds}=0,\label{weyl-geod-1}\eea or after an appropriate redefinition of the affine parameter, $ds\rightarrow e^{-\vphi/2}d\tau$: 

\bea \frac{d^2x^\mu}{d\tau^2}+\Gamma^\mu_{\sigma\lambda}\frac{dx^\sigma}{d\tau}\frac{dx^\lambda}{d\tau}=0,\label{weyl-geod}\eea under the Weyl rescalings \eqref{weyl-t}. In the last equation the affine parameter $d\tau$ is not transformed by the conformal transformation.

%-------------------------------------------------------------------

\subsection{Gauge freedom: the GR gauge}\label{subsect-g-free}

It is apparent from \eqref{weyl-inv-theor} that there is not an independent equation of motion for the gauge field $\vphi$.  As a matter of fact, the Einstein's equation in \eqref{weyl-inv-theor} and the conservation equation \eqref{conserv-eq}, are the only independent motion equations of the above Weyl-invariant theory of gravity. This property is the inevitable consequence of invariance of the theory under the Weyl rescalings \eqref{weyl-t}. Hence, in addition to the four degrees of freedom to make diffeomorphisms, there is an additional degree of freedom to make conformal transformations. In other words: we are free to choose any function $\vphi=\vphi(x)$ we want provided it is continuous and (at least twice) differentiable.

A particularly simple and outstanding gauge is the one given by the choice: $\vphi=\vphi_0$, where $\vphi_0$ is a constant. Under this choice the affine connection of Weyl-integrable spaces is transformed into the Christoffel symbols of the metric and the metricity condition of WIG is transformed into the Riemann metricity property: the metric tensor is covariantly constant. This means, in turn, that the Weyl-integrable spaces are transformed into spaces with (pseudo)Riemann geometric structure, as those in GR. In correspondence with this choice of gauge, the motion equation in \eqref{weyl-inv-theor} transforms into the Einstein's equation of general relativity: $$G_{\mu\nu}=8\pi G_N\left(T^{(m)}_{\mu\nu}-\Lambda g_{\mu\nu}\right),$$ with $$8\pi G_N\equiv\frac{1}{M^2_\text{Pl} e^{\vphi_0}},\;\;\Lambda\equiv\lambda e^{2\vphi_0}.$$ The Einstein's equation is supplemented with the standard conservation equation: $\nabla^\mu T^{(m)}_{\mu\nu}=0$. 

In other words, the choice $\vphi=\vphi_0$ corresponds to plain general relativity, and we call this as the GR gauge. Hence, GR is one of the infinitely many physically equivalent representations of the Weyl-invariant laws of gravity \eqref{weyl-inv-theor}. This is why we call the above Weyl-invariant theory of gravity as ``extended general relativity'' (EGR). 

There can be, of course, infinitely many possible choices of the gauge field $\vphi$, with each one of the choices leading to a potential description of the gravitational laws according to EGR. Measurable quantities are independent of the above choice since these are necessarily invariant under the Weyl rescalings \eqref{weyl-t}. For instance, the gauge invariant measure of the curvature scalar is defined by, $R_*:=e^{-\vphi}R^{(w)}$, while the gauge-invariant measure of spacetime separations (line-element squared) is $ds^2_*:=e^\vphi ds^2$. Other Weyl-invariant quantities are: $$e^{-2\vphi}R^{(w)}_{\mu\nu}R_{(w)}^{\mu\nu},\;e^{-3\vphi}R^{(w)}_{\mu\nu\tau\lambda}R_{(w)}^{\mu\nu\tau\lambda},$$ among others. Let us focus, for illustration, in the Weyl-invariant Kretschmann scalar: $$K_*:=e^{-3\vphi}R^{(w)}_{\mu\nu\tau\lambda}R_{(w)}^{\mu\nu\tau\lambda}.$$ In the GR gauge, for instance, $$K_*=e^{-3\vphi_0}R_{\mu\nu\tau\lambda}R^{\mu\nu\tau\lambda},$$ where $R_{\mu\nu\tau\lambda}$ is the standard (GR) Riemann-Christoffel curvature tensor. Hence, we have that: 

\bea K_\text{GR}=R_{\mu\nu\tau\lambda}R^{\mu\nu\tau\lambda}=e^{-3(\vphi-\vphi_0)}R^{(w)}_{\mu\nu\tau\lambda}R_{(w)}^{\mu\nu\tau\lambda},\label{gr-kretschmann}\eea where $K_\text{GR}$ is the standard (Riemannian) Kretschmann invariant. The above equation will be useful in section \ref{sect-sing}, when we discuss on the singularity issue.

%---------------------------------------------------------------------------

\subsection{Coupling the SMP in a Weyl-invariant way}\label{subsect-weyl-smp}

In order to incorporate the standard model of elementary particles in a Weyl-invariant way into the theory \eqref{weyl-inv-theor}, the EW Lagrangian for the Higgs field could be written as:

\bea {\cal L}_H=-\frac{1}{2}|DH|^2-\frac{\lambda'}{2}\left[|H|^2-v^2e^\vphi\right]^2,\label{higgs-lag}\eea with $|H|^2\equiv H^\dag H$ and $|DH|^2\equiv g^{\mu\nu}(D_\mu H)^\dag(D_\nu H)$, where the gauge covariant derivative: 

\bea D_\mu H\equiv(D^*_\mu-\frac{1}{2}\der_\mu\vphi)H.\label{gauge-der}\eea Here $$D^*_\mu H\equiv\left(\der_\mu+\frac{i}{2}g W_\mu^k\sigma^k+\frac{i}{2}g'B_\mu\right)H,$$ is the gauge covariant derivative of the EW theory with $W_\mu^k=(W_\mu^\pm,W_\mu^0)$-- the SU(2) bosons, $B_\mu$-- the U(1) boson, $\sigma^k$-- the Pauli matrices and $(g,g')$-- the gauge couplings. Notice that, under the Weyl rescalings \eqref{weyl-t}: $$H\rightarrow\Omega\,H,\;\left(W_\mu^k,B_\mu\right)\rightarrow\left(W_\mu^k,B_\mu\right),$$ while the enlargement of the gauge covariant derivative \eqref{gauge-der} to include the derivative of the Weyl gauge scalar, allows the corresponding operator to preserve, also, the Weyl symmetry. This entails that, $D_\mu\rightarrow D_\mu$, and that, besides, the gauge derivative commutes with the conformal factor: $[D_\mu,\Omega]=0$, so that under \eqref{weyl-t}, $$D_\mu H\rightarrow\Omega D_\mu H\;\Rightarrow\;|DH|^2\rightarrow\Omega^4|DH|^2,$$ leading to: ${\cal L}_H\rightarrow\Omega^4{\cal L}_H$, i. e., the EW Higgs field action piece, $S_H=\int d^4x\sqrt{|g|}{\cal L}_H$, is not transformed by the Weyl rescalings. Above we have assumed that the ``bare'' constants $g$, $g'$, $\lambda'$ and $v$ are not transformed by \eqref{weyl-t} (see the following section for a discussion on the transformation properties of the constants of nature). 

In the SMP supported by the Weyl-invariant Lagrangian \eqref{higgs-lag}, after ``symmetry breaking'' the Higgs acquires a point-dependent VEV; $|H|=v\,e^{\vphi/2}$, so that the gauge bosons and fermions of the SMP acquire point-dependent masses: $m_p(\vphi)=g_p\,v\,e^{\vphi/2}$, where $g_p$ is some gauge coupling. The resulting theory of gravity and of the SMP preserves the Weyl symmetry even after symmetry breaking.\footnote{The Weyl invariance of the action for fermion fields in curved spacetime is shown in Ref. \cite{cheng_prl_1988} (see also the appendix B of Ref. \cite{sc_inv_quiros}).} This means that Weyl invariance may be an actual symmetry of the laws of physics in our present universe.

%%%%%%%%%%%%%%%%%%%%%%%%%%%%%%%%%%%%%%%%%%%%%%%%%%%%%%%%%%%%%%%%%%%%%%%%%%%%%%%%%%%%%%%%%%%%%%%%%%

\section{The constants of nature and their transformation properties}\label{sect-const-nature}

An aspect of the theory of gravity \eqref{weyl-inv-theor} that is worthy of noticing, is related with the transformation properties of dimensionful constants under the Weyl rescalings \eqref{weyl-t}. In the bibliography (see, for instance, \cite{dicke-1962, faraoni_prd_2007}) one founds arguments that point to dimensionless constants as being unchanged by the Weyl rescalings, while dimensionful constants should be necessarily transformed under \eqref{weyl-t}. Although at first sign this might appear as a natural statement, in the same cited references one founds dimensionful constants, like the Planck constant $\hbar$, the speed of light $c$ and the electric charge of the electron $e$, that are assumed to be unchanged by the Weyl rescalings. In this regard, let us assume that the (dimensionful) constant mass of a given particle, say $m_p$, transforms like \cite{dicke-1962, faraoni_prd_2007}: $m_p\rightarrow\Omega\,m_p$, under the Weyl rescalings. Since, according to \eqref{weyl-t}: $e^{\vphi/2}\rightarrow\Omega\,e^{\vphi/2}$, then one may conclude that the mass of a particle -- assumed to be a constant from the start -- should be a point-dependent quantity as well: $m_p\propto e^{\vphi/2}$. The inconsistency arises from the fact that Riemannian geometry -- usually assumed to be the underlying geometrical structure of spacetime -- does not admit varying constants since the affine law of the Riemannian manifolds does not admit the length of vectors to vary under parallel transport. Hence, any assumption on the transformation properties of the constants of the theory under the Weyl rescalings, can be either an independent postulate of the theory or a meaningless statement otherwise. Following this reasoning line we want to criticize a statement frequently found in the bibliography on scale invariance: ``Weyl symmetry does not allow any dimensionful parameters in the action.'' Such a statement were correct if implicitly assume Riemann geometry to govern the affine properties of spacetime, i. e., if these parameters were not supposed to vary from point to point in spacetime, but not in general. 

It is important to notice that the fundamental dimensionful constants such as the Plack mass, are included in the action \eqref{weyl-inv-theor} from the start. In this regard, the theory \eqref{weyl-inv-theor} is to be considered as less fundamental than, for instance, the ones in \cite{smolin_npb_1979, zee_prd_1981, cheng_prl_1988, sc_inv_prester, sc_inv_bars, sc_inv_bars_1, sc_inv_bars_2, sc_inv_carrasco, sc_inv_jackiw, sc_inv_alpha, sc_inv_alpha_2, sc_inv_alpha_3, sc_inv_alpha_4}. In particular, the emergence of fundamental scales can not be addressed within this setup \cite{sc_inv_quiros}. Besides, the action (\ref{weyl-inv-theor}) differs from the one in \cite{sc_inv_prester, sc_inv_bars, sc_inv_bars_1, sc_inv_bars_2, sc_inv_carrasco, sc_inv_jackiw, sc_inv_alpha, sc_inv_alpha_2, sc_inv_alpha_3, sc_inv_alpha_4}, in that the underlying geometric structure is WIG, and $\vphi$ is no longer another singlet scalar field but it is just the Weyl gauge field of WIG, i. e., the $\vphi$-kinetic energy term is already included in the Weyl-integrable curvature scalar:

\bea R^{(w)}=R-\frac{3}{2}(\der\vphi)^2-3\Box\vphi,\label{curv-scalar-w}\eea where in the RHS of this equation the given quantities and operators coincide with their Riemannian definitions in terms of the Christoffel symbols of the metric.

The main consistency assumption of the theory \eqref{weyl-inv-theor} (see Ref. \cite{sc_inv_quiros}) is that only the fields may be transformed by the Weyl rescalings \eqref{weyl-t}. In this regard, for instance, the BD-type scalars like in \eqref{deser-action} are transformed in the same way as the (square root of the logarithm of the) gauge scalar: $\phi\rightarrow\Omega\,\phi$, while for vectors, $V^\mu\rightarrow\Omega^2\,V^\mu$ ($V_\mu\rightarrow\,V_\mu$). For the fermion fields $\psi$ the transformation under \eqref{weyl-t} reads: $\psi\rightarrow\Omega^{3/2}\,\psi$, etc. The actual constants are not transformed by the Weyl transformations \eqref{weyl-t}. This is their distinguishing feature. However, if the given dimensionful constant is multiplied by an appropriate power of some field, the resulting quantity is not a constant any more but a field. Take as an example the varying Planck mass squared: $M^2_\text{Pl}(\vphi)=M^2_\text{Pl} e^\vphi$, where under (\ref{weyl-t}), $M^2_\text{Pl}(\vphi)\rightarrow\Omega^2 M^2_\text{Pl}(\vphi)$, then the resulting point-dependent Plack mass squared does actually transform under the Weyl rescalings, just as a scalar field does. In this case Weyl invariance might be preserved if consider terms like $M^2_\text{pl}(\vphi)\,R^{(w)}$, or $M^4_\text{Pl}(\vphi)$, in the action. The occurrence of such varying ``constants'' in a WIG-based Weyl-invariant theory of gravity is the natural consequence of the variations of the length of geometrical objects (vectors, tensors, etc.) during parallel transport in manifolds with WIG structure.

The price to pay for allowing point-dependent fundamental constants in the scale-invariant action from the start, is to renounce to ``natural'' generation of those fundamental constants by means of symmetry breaking arguments. In the present paper, following Ref. \cite{sc_inv_quiros}, we shall call the actual constants like $\hbar$, $c$, $M_\text{Pl}$, $v$, etc., as ``bare'' constants, in contrast to point-dependent ``constants'' which are obtained as a combination of a bare constant and some power of the Weyl gauge scalar. In the action \eqref{weyl-inv-theor}, as well as in \eqref{higgs-lag}, we have included the point-dependent Planck mass $M_\text{Pl}$ and EW mass parameter $v$, respectively: $M^2_\text{Pl}(\vphi)=M^2_\text{Pl}\,e^\vphi$, $v^2(\vphi)=v^2\,e^\vphi$, where the constants carry appropriate units, while the gauge scalar $\vphi$ is dimensionless. As already said, these point-dependent ``constants'' arise naturally in spacetimes whose affine geometrical structure is governed by Weyl integrable geometry or any other modification of Riemann geometry, which allows the lengths of vectors -- and the associated units of measure -- to vary from point to point in spacetime.

%%%%%%%%%%%%%%%%%%%%%%%%%%%%%%%%%%%%%%%%%%%%%%%%%%%%%%%%%%%%%%%%%%%%%%%%

\section{Cosmology in the extended general relativity}\label{sect-cosmo}

Let us to search for cosmological solutions of the motion equations in \eqref{weyl-inv-theor}. As before, here we consider a FRW background spacetime with flat spatial sections, with line-element: $ds^2=-dt^2+a^2(t)\delta_{ik}dx^idx^k$. For simplicity here we take the case with $\lambda=0$, so that the second term in the RHS of \eqref{weyl-cosmo-moteq} vanishes. The case with $\lambda\neq 0$ will be considered separately. The Friedmann equation in \eqref{weyl-inv-theor} and the conservation equation \eqref{conserv-eq} read,

\bea &&3\left(H+\frac{\dot\vphi}{2}\right)^2=\frac{\rho^{(w)}_m}{M^2_\text{Pl}},\nonumber\\
&&\dot\rho^{(w)}_m+3\gamma\left(H+\frac{\dot\vphi}{2}\right)\rho^{(w)}_m=0,\label{weyl-moteq}\eea or, after integrating the second equation above:

\bea 3\left(H+\frac{\dot\vphi}{2}\right)^2=\frac{M^4}{M^2_\text{Pl}}\frac{e^{-3\gamma\vphi/2}}{a^{3\gamma}},\label{weyl-cosmo-moteq}\eea where $M^4$ is an integration constant. In the above equations the matter contribution have been assumed in the form of a perfect barotropic fluid with energy density, $\rho^{(w)}_m\equiv e^{-\vphi}\rho_m$, and pressure, $p^{(w)}_m$, which obey the equation of state: $p^{(w)}_m=(\gamma-1)\rho^{(w)}_m$ ($\gamma$ is the barotropic index). 

Let us introduce the variable, $Z\equiv\ln a+\vphi/2$, so that \eqref{weyl-cosmo-moteq} reads: $$3e^{3\gamma Z}\dot Z^2=\frac{M^4}{M^2_\text{Pl}}.$$ Straightforward integration of the above equation yields to:

\bea a(t)e^{\vphi(t)}=\alpha^\frac{2}{3\gamma}\left(t-t_0\right)^\frac{2}{3\gamma},\label{sol-1}\eea where $t_0$ is an (rescaled) integration constant and $$\alpha\equiv\frac{3\gamma}{2}\sqrt\frac{M^4}{3M^2_\text{Pl}}.$$ 

Given the gauge freedom inherent in the Weyl invariant theory of gravity \eqref{weyl-inv-theor}, only the combination $ae^\vphi$ can be determined by the motion equations. We are free to choose either any $a(t)$ or any $\phi(t)$ we want. Assume, for instance, the following cosmic dynamics: $a(t)=a_0(t-t_0)^n$. In this case the dynamics of the scalar field is given by: $$\vphi(t)=\vphi_0+\left(\frac{2}{3\gamma}-n\right)\ln(t-t_0),$$ where $\vphi_0\equiv\ln(\alpha^{2/3\gamma}/a_0)$. Notice that the choice, $n=2/3\gamma$, leads to the GR gauge where: $\vphi=\vphi_0$ and $a(t)=a_0(t-t_0)^{2/3\gamma}$. 

Another interesting situation is when the background fluid is vacuum: $\gamma=0$. In this case, integration of \eqref{weyl-cosmo-moteq} yields: $$a(t)e^{\vphi(t)/2}=C_0\exp\left(\sqrt\frac{M^4}{3M^2_\text{Pl}}\,t\right),$$ where $C_0$ is (the exponent of) an integration constant. Hence, the choice of the GR gauge where $\vphi=\vphi_0$, leads to de Sitter expansion: $$a(t)\propto e^{\sqrt{\Lambda/3}\,t},\;\;\Lambda=\frac{M^4}{M^2_\text{Pl}}.$$ 

Another particularly interesting gauge is when, $\vphi(t)=\vphi_0+2\sqrt{M^4/3M^2_\text{Pl}}\,t$. In this gauge, $a=C_0 e^{-\vphi_0/2}$, is a constant so that we get a static universe. This means that de Sitter expansion in GR where the space is taken to be Riemannian, is physically equivalent to a static universe where the affinity of the space is Weyl integrable. 

How it can be that the same experimental data that supports Riemannian-de Sitter expansion can serve as experimental evidence for a Weyl-integrable static universe? The prompt answer is quite simple: While in the Riemannian de Sitter space the measured redshift is due to the expansion of the universe, in the Weyl-integrable space the same redshift is due to variation of the units of measure during the cosmic evolution. Imagine that as a result of certain atomic transition a photon with frequency, $\hbar\omega\propto\Delta m_\text{Atom}\propto\exp(\vphi(t_p)/2)$, is emitted at some time $t_p$ in the past in a distant location, in a Weyl-integrable static space. The energy of the photon ($\hbar\omega$) when received at some latter time $t_0$, is not enough to excite the same atomic transition in a similar atom: $\Delta m^0_\text{Atom}\propto\exp(\vphi(t_0)/2)$. The measured redshift would be, $$\frac{\Delta m^0_\text{Atom}-\Delta m_\text{Atom}}{\hbar}\approx\sqrt\frac{M^4}{3M^2_\text{Pl}}\frac{(t_0-t_p)}{\hbar},$$ where we have assumed that $t_0, t_P\ll\sqrt{3M^2_\text{Pl}/M^4}$, i. e. we are considering time periods much more shorter that the lifetime of the universe $\sim H^{-1}_0\approx 10^{18}$ sec. 

Hence, even if the universe is static, in a Weyl-integrable space, due to the point-dependent property of the units of measure, there is room for the redshift. This point will be discussed in detail in section \ref{subsect-discu-weyl} below.

%----------------------------------------------------

\subsection{Background without matter and $\lambda\neq 0$}

In this particular case we take $\rho_m^{(w)}=0$, and the motion equation in \eqref{weyl-inv-theor} reads:

\bea H+\frac{\dot\vphi}{2}=\pm\sqrt\frac{\lambda}{3M^2_\text{Pl}}\,e^{\vphi/2}.\label{lambda-sol}\eea For definiteness let us choose the positive branch and the particular case where $H=H_0$, i. e., again the de Sitter expansion. We get: $$\frac{\dot\vphi}{2}=\sqrt\frac{\lambda}{3M^2_\text{Pl}}\,e^{\vphi/2}-H_0,$$ which, after straightforward integration yields; $$e^{\vphi(t)/2}=\frac{H_0}{C_0H_0\,e^{H_0\,t}+\sqrt\frac{\lambda}{3M^2_\text{Pl}}},$$ where $C_0$ is (the exponent of) an integration constant. It is seen that in the formal limit $t\rightarrow-\infty$, the Weyl gauge scalar, $$\vphi\approx\ln\left(\frac{3M^2_\text{Pl}}{\lambda}\,H_0^2\right),$$ is basically a constant so that the GR gauge is approached in this limit. This is an example of a cosmic dynamics that continuously joints the GR gauge with another (non-GR) gauge.

%%%%%%%%%%%%%%%%%%%%%%%%%%%%%%%%%%%%%%%%%%%%%%%%%%%%%%%%%%%%%%%%%%%%%%%%%%%%%%%%

\section{Extended general relativity and the singularity issue}\label{sect-sing}

The apparent paradox that arises when comparing solutions with a spacetime singularity in one frame with the same solutions in a conformal frame where the singularity may be absent, has been discussed several times without a clear resolution \cite{faraoni_prd_2007, quiros_grg_2013, kaloper_prd_1998, quiros_prd_2000}. In this regard we have to say that the issue has been discussed within the framework of the Brans-Dicke theory -- in its JF and EF representations -- where there is no room for conformal invariance of the laws of gravity and, hence, the assumed physical equivalence of the conformal frames may be unjustified. In Ref. \cite{faraoni_prd_2007}, for instance, it is said that: ``If, following Dicke \cite{dicke-1962}, the Jordan and Einstein frames are equivalent, singularities occur in the Einstein frame if and only if they occur in the Jordan frame.'' Then it is apparently demonstrated that given a bigbang singularity in the JF, in the EF the singularity will be still there. The demonstration relies on the assumption of running units in the Einstein's frame. This assumption could be correct if suppose, besides, that the geometric structure of the spacetime is Weyl-integrable, but not in a (pseudo)Riemannian manifold where there is no room for varying length of vectors under parallel transport.\footnote{In Ref. \cite{quiros_prd_2000} the issue was discussed under the assumption that the underlying geometrical structure of the background spacetime is WIG, but the action considered (BD action) is itself not conformal invariant, so that the conformal frames are not actually physically equivalent.} Although (in principle) one may be free to assume the behavior of the units of measurement as an independent postulate of a given theory, this is not without costs. In particular this would entail that the masses of particles may be point-dependent if the units of measure are so, and that certain effects usually adscribed to the curvature as, for instance, the redshift, may be partly due to the curvature and partly due to the assumed point-dependent property of the measurement units. Hence, the assumption is at the cost of assigning gravitational effects to the units of measurement (and to the measurement process) themselves. 

As we shall show below, such kind of statement conditioning the unavoidance of spacetime singularities in the different conformally related frames to their physical equivalence, is wrong in general, even if the conformal frames (or gauges) are actually physically equivalent. In order to show this we shall consider the Weyl invariant theory \eqref{weyl-inv-theor}, i. e. the so called EGR (see sections \ref{sect-egr}, \ref{sect-const-nature} and \ref{sect-cosmo} above). In this theory the different conformal frames or gauges, are actually physically equivalent given that, not only the action of the theory and the derived motion equations are invariant under Weyl rescalings \eqref{weyl-t}, but, also the assumed WIG-structure of the background spacetimes, warrants that the geodesics of the geometry are conformal invariant. This, in turn, eliminates the possibility of the anomalous coupling of matter in Weyl-invariant theories where the underlying structure of the background spacetimes is Riemannian (see section \ref{sect-scale-inv} and, specially, subsection \ref{subsect-anomalous-c}). 

Since the quantities that have the physical meaning are those which are not only invariant under general coordinate transformations but, at the same time, are also invariant under the Weyl rescalings \eqref{weyl-t} -- here we call these as gauge invariant quantities -- our discussion will rely exclusively on these gauge invariant quantities. Take, for instance, the gauge-invariant measure of spacetime separations: $ds_*^2=e^\vphi g_{\mu\nu}dx^\mu dx^\nu$. Consider the GR-gauge, where $\vphi=\vphi_0$ (for simplicity let us take $\vphi_0=0$). We have that, $ds_*^2=ds^2_\text{GR}=g^\text{GR}_{\mu\nu}dx^\mu dx^\nu.$ This entails the following relationship between the GR metric tensor and a given conformal metric:

\bea g_{\mu\nu}=e^{-\vphi}g_{\mu\nu}^\text{GR}.\label{metric-rel}\eea Let us further assume, for definiteness, the spherically symmetric Schwarzschild GR vacuum metric, $$ds^2_\text{GR}=-\left(1-\frac{2m}{r}\right)dt^2+\left(1-\frac{2m}{r}\right)^{-1}dr^2+r^2d\Omega^2,$$ where $d\Omega^2\equiv d\theta^2+\sin^2\theta d\phi^2$. Since we can freely choose the gauge scalar, let us set:

\bea \vphi(r)=2q\ln\left(1-\frac{2m}{r}\right),\label{vphi-r}\eea where $q$ is an arbitrary constant that parametrizes our choice for the gauge scalar. Hence, we have a set of physically equivalent WIG descriptions of the laws of gravity \eqref{weyl-inv-theor} given by: $\{({\cal M},g^q_{\mu\nu},\vphi_q):\,q\geq 0\}$, where ${\cal M}$ represents the spacetime manifold. The values $q<0$ are not considered since, as shown in Ref. \cite{quiros_prd_2000}, in this case one gets a set of spacetimes with naked singularities. Notice that general relativity is included in the above equivalence class as the representation specified by the choice, $q=0$. For the $q$-th representation the metric reads:

\bea &&ds^2=-\left(1-\frac{2m}{r}\right)^{1-2q}dt^2\nonumber\\
&&\;\;\;\;\;\;\;\;\;\;\;+\left(1-\frac{2m}{r}\right)^{-1-2q}dr^2+\rho^2d\Omega^2,\label{ds-q}\eea where, $$\rho(r)=\frac{r}{(1-2m/r)^q},$$ is the proper radial coordinate. Since $\rho$ could be non-negative, then: $2m\leq r<\infty$. Besides, the proper radial coordinate is a minimum at $r_\text{min}=2(1+q)m$, where 

\bea \rho_\text{min}=\frac{(1+q)^{1+q}}{q^q}\,2m.\label{min-rho}\eea 

In order to discuss on the occurrence of spacetime singularities it is useful to write the relationship between the GR Kretschmann scalar, $K_\text{GR}=R_{\mu\nu\tau\lambda}R^{\mu\nu\tau\lambda}=48m^2/r^6$, and the one for the conformal WIG representation, $K_{(w)}=R^{(w)}_{\mu\nu\tau\lambda}R_{(w)}^{\mu\nu\tau\lambda}$, given in \eqref{gr-kretschmann} with $\vphi_0=0$: 

\bea K_{(w)}=e^{3\vphi}K_\text{GR}=\frac{48m^2}{r^6}\left(1-\frac{2m}{r}\right)^{6q}=\frac{48m^2}{\rho^6}.\label{kretsh-rel}\eea 

A class of spacetimes with two asymptotically flat spatial infinities is obtained: one at $r\rightarrow\infty$ and the other one at $r\rightarrow 2m$ where, in both cases, $\rho\rightarrow\infty$ while $K_{(w)}\rightarrow 0$. These spatial infinities are joined by a throat with minimum value of the proper radial coordinate, $\rho_\text{min}$, given by \eqref{min-rho}, where the curvature is a maximum: $$K_{(w)}^\text{max}=\frac{3q^{6q}}{4(1+q)^{6(1+q)}m^4}.$$ A similar result was discussed in Ref. \cite{quiros_prd_2000}, however, as already mentioned, in that reference the BD theory was considered so that the analysis performed was not based on the gauge invariants like in the present section. Other discussions on the singularity issue can be found \cite{faraoni_prd_2007, kaloper_prd_1998} where, once again, due to the lack of invariance under the Weyl rescalings \eqref{weyl-t}, the analysis could not be based on the gauge invariants.\footnote{It has been shown in Ref. \cite{veermae} that the existence of wormhole vacua in conformal (Weyl-invariant) theories of gravity is a completely general and unavoidable feature of conformal gravity.}

The result we have just obtained is a clear example of the fact that a given spacetime singularity existing in one or in several physically equivalent frames, may be safely avoided in other physically equivalent representations. In the present case, a Schwarzschild black hole with a singularity at $r=0$, that is enclosed by an event horizon at $r=2m$, occurring in the GR gauge, in a class of conformal, physically equivalent gauges, $\{({\cal M},g^q_{\mu\nu},\vphi_q):\,q>0\}$, is replaced by wormhole spacetimes which are free of singularities.

%%%%%%%%%%%%%%%%%%%%%%%%%%%%%%%%%%%%%%%%%%%%%%%%%%%%%%%%%%%%%%%%%%%%%%%%%%%%

\section{Discussion}\label{sect-discussion}

There are two main results that, despite being intimately related, we want to discuss separately. The first result we want to discuss is about the meaning of conformal invariance or Weyl invariance in connection with the conformal transformations' issue, while the second result is related with the unavoidable physical consequences of Weyl invariance.

%------------------------------------------------------------------------------------------------------------------

\subsection{Conformal transformations and physical equivalence of the conformal frames}\label{subsect-discu-conf-t}

When discussing on physical equivalence of different representations/frames/gauges of a given theory, one almost immediately thinks on general relativity or on the gauge theories of the interactions, where the different coordinate frames in the former, or the different gauges in the latter, carry no physical meaning at all. As a matter of fact the different coordinate frames in which the laws of gravity can be formulated according to GR, are physically equivalent, so that from the physical standpoint none of the coordinate frames is preferred over the others. The same is true of any of the different gauges in which the given gauge theory may be formulated. Hence, it is difficult to understand how one can discuss on physical equivalence of the conformal frames in which the scalar-tensor theories can be formulated, when these theories do not actually embody conformal invariance of the laws of gravity. What can be understood by physical equivalence in such cases? Physical equivalence necessarily entails a underlying symmetry which, in turn, means some freedom in the choice of given field variable(s). When the conformal transformation is considered, in addition to the four degrees of freedom to make diffeomorphisms, an extra degree of freedom arises in connection with the conformal symmetry. But in the Brans-Dicke theory, as in the scalar-tensor theories in general, there is not any additional degree of freedom in connection with the assumed conformal invariance of the gravitational laws!

Regardless of this, in almost every work on the issue where physical equivalence of the conformal frames is invoked, it is mentioned that the laws of physics must be invariant under units' transformations understood as conformal transformations of the metric (see, for instance, Ref. \cite{dicke-1962, faraoni_prd_2007, sotiriou_etall_ijmpd_2008, quiros_prd_2000}). But, the mere existence of the different conformal frames, like the JF and the EF, is an evidence against invariance of the Brans-Dicke (also of general scalar-tensor) laws of gravity under the conformal transformations of the metric, where by ``laws of gravity'' we understand the action principle together with the derived equations of motion. This is why we think that the theoretical basis for the conformal transformations' issue is doubtful.

In the present paper, in order to discuss on physical equivalence in connection with conformal invariance, we have chosen an example of an actually conformal invariant theory of gravity \eqref{weyl-inv-theor}, where not only the equations of motion (and the action) are invariant under the Weyl rescalings \eqref{weyl-t}, but also the geodesics of the geometry are unchanged by \eqref{weyl-t}. This is a very important point to consider since there are many works on local scale-invariant theories of gravity (see, for instance, Refs. \cite{sc_inv_padilla, sc_inv_bars, sc_inv_bars_1, sc_inv_bars_2, sc_inv_alpha, sc_inv_alpha_2, sc_inv_alpha_3, sc_inv_alpha_4}), where the geometrical aspect is not adequately discussed. Moreover, it is implicitly assumed that the underlying geometrical structure of the background spacetime is Riemann geometry. But, as shown in subsection \ref{subsect-anomalous-c}, this geometrical structure is not well suited to address Weyl invariance, since the geodesics of the metric in Riemann geometry are indeed transformed by the conformal transformations.

%------------------------------------------------------------------------------

\subsection{Physical implications of Weyl invariance}\label{subsect-discu-weyl}

As discussed above, Weyl invariance of the physical laws -- in particular of the gravitational laws -- implies invariance of the action of the theory and, consequently, of the resulting equations of motion, under Weyl rescalings \eqref{weyl-t}: $$g_{\mu\nu}\rightarrow\Omega^{-2}g_{\mu\nu},\;\;\vphi\rightarrow\vphi+2\ln\Omega.$$ Additionally, it also requires invariance of the geodesics of the geometry under these transformations since, otherwise, an issue with anomalous coupling of matter inevitably arises (see subsection \ref{subsect-anomalous-c}). This is achieved only if substitute the usually assumed Riemann geometrical structure of the background spacetime, by the Weyl-integrable geometry, where the units of measure are allowed to depend on the spacetime point. We end up with a theory where Weyl invariance is explicit in the action/field equations \eqref{weyl-inv-theor}: 

\bea &&S=\int d^4x\sqrt{|g|}\left[\frac{M^2_\text{Pl}e^\vphi}{2} R^{(w)}-\lambda e^{2\vphi}+{\cal L}_m\right],\nonumber\\
&&G^{(w)}_{\mu\nu}=\frac{e^{-\vphi}}{M^2_\text{Pl}}\,T^{(m)}_{\mu\nu}-\frac{\lambda e^\vphi}{M^2_\text{Pl}}g_{\mu\nu},\nonumber\eea and where the geometric quantities and operators are defined in spacetimes with Weyl-integrable affinity (also explicit in the action/motion equations where these quantities are identified by a supra/sub-index $(w)$). This theory, called here as EGR, represents an actual example of a fully Weyl-invariant theory, where by ``fully'' we understand that there are not problems with a consistent coupling of matter degrees of freedom (other than radiation). 

In such kind of theory that actually embodies physical equivalence of the different conformal frames/gauges, the gauge scalar $\vphi$ -- or one of the components of the metric -- can be chosen at will given the additional degree of freedom associated with Weyl invariance. The consequences of this gauge freedom are actually astonishing: The observational evidence for a spherically symmetric (Schwarzschild) black hole that is recorded by observers living in the GR-gauge, is interpreted by observers living in a conformally related gauge as a wormhole -- without singularity -- joining two asymptotically flat spatial regions (see section \ref{sect-sing}).  Another example of this gauge freedom may be found within the cosmological setting (see section \ref{sect-cosmo}). It has been shown that a de Sitter space in the GR-gauge is physically equivalent to a Weyl-integrable static universe in a properly chosen conformal gauge. In this case, while in the GR gauge the redshift data suggests exponentially fast expansion, in the given WIG conformal frame although the metric is flat (the universe is static), the non-vanishing Weyl gauge scalar also contributes towards the curvature of spacetime, resulting in a redshift effect associated with the point-dependent property of the units of measure. Below we shall explain what happens in more detail.

%-----------------------------------------------

\subsubsection{On WIG geodesics and the redshift}

One of the indisputable assumptions of the conformal transformations procedure is that the co-variant components of the vector potential $A_\mu$ are not transformed by the conformal transformation in \eqref{weyl-t}. Hence, consistency of the procedure requires that the co-variant components of any other vector, be it the 4-wavevector $k_\mu$ of a photon, or the 4-momentum of a particle $p_\mu$, would not be transformed by the Weyl rescalings: $(A_\mu,k_\mu,p_\mu,...)\rightarrow(A_\mu,k_\mu,p_\mu,...)$. This means that under the aforementioned transformations, $$(A^\mu,k^\mu,p^\mu,...)\rightarrow\Omega^2(A^\mu,k^\mu,p^\mu,...).$$ Hence, the Weyl invariant WIG geodesic equations for particles with point dependent mass $m(\vphi)=m_0 e^{\vphi/2}$ and 4-momentum, $$p^\mu=m(\vphi)\frac{dx^\mu}{ids},$$ read (compare with equation \eqref{weyl-geod-1}): 

\bea \frac{dp^\mu}{ds}+\Gamma^\mu_{\sigma\lambda}\frac{dx^\sigma}{ds}p^\lambda-\der_\lambda\vphi\frac{dx^\lambda}{ds}p^\mu=0.\label{4-p-geod}\eea The 4-wavevector $k^\mu=(w,{\bf k})$ of a photon obeys the same geodesic equation after simultaneously replacing $p^\mu\rightarrow k^\mu$, and the line element by the differential of an affine parameter $\sigma$ along the photon path: $ds\rightarrow d\sigma$. We recall that, under the Weyl rescalings \eqref{weyl-t}, $(ds,d\sigma)\rightarrow\Omega^{-1}(ds,d\sigma)$. The subtle point here is that, if in the above geodesic equations replace the WIG affine connection, $\Gamma^\mu_{\sigma\lambda}$, by its expression through the Christoffel symbols of the metric, $\{^\mu_{\sigma\lambda}\}$ in \eqref{weyl-aff-c}, for the geodesic of a particle with 4-momentum we get: $$\frac{dp^\mu}{ds}+\{^\mu_{\sigma\lambda}\}\frac{dx^\sigma}{ds}p^\lambda-\frac{1}{2}g_{\sigma\lambda}\frac{dx^\sigma}{ds}p^\lambda\der^\mu\vphi=0,$$ while for a photon, since $g_{\sigma\lambda}dx^\sigma dx^\lambda=0$, the geodesic: 

\bea \frac{dk^\mu}{d\sigma}+\{^\mu_{\sigma\lambda}\}\frac{dx^\sigma}{d\sigma}k^\lambda=0,\label{photon-geod}\eea coincides with the geodesic of a photon in a Riemannian spacetime. This is another way to show that the geodesics of massless particles are not affected by the conformal transformations of the metric. In other words, the photon is blind to the affinity of the spacetime: it does not differentiate a Riemannian space from a Weyl-integrable one.

Consider next a static FRW space (the scale factor is a constant), so that the Christoffel symbols of the metric all vanish. From the geodesic equation for particles with the mass it follows that the 4-momentum changes along the geodesic, meanwhile, from the photon's geodesic equation it follows that the 4-wavevector is unchanged under parallel transport along the geodesic. Let us assume that the (rest) energy of a given atomic transition in a given spacetime point $P$, is given by: $\Delta m(\vphi(P))=\Delta m_0\exp(\vphi(P)/2)$. As a result of the transition a photon of energy $\hbar\omega_P=\Delta m(\vphi(P))$ is emitted. Suppose now that an identical atom that is placed at another spacetime point $P_0$, also emits a photon as a result of the same transition. The energy of the emitted photon is given by: $\hbar\omega_{P_0}=\Delta m_0\exp(\vphi(P_0)/2).$ Let us assume that the photon emitted at $P$ reaches $P_0$ so that its energy can be compared with the energy of the photon emitted at $P_0$. Since, according to \eqref{photon-geod}, in a static universe the energy carried by the photon is unchanged along the photon's path, the relative magnitude of the redshift in the case discussed here, is given by: $$\frac{\Delta\omega}{\omega_P}=\frac{\left(\omega_{P_0}-\omega_P\right)}{\omega_{P_0}}=1-\exp\left(\frac{\vphi(P)-\vphi(P_0)}{2}\right).$$ Hence, the occurrence of a non-vanishing redshift in a static WIG universe is due to the fact that the masses of particles -- including atoms, etc. -- that emit photons, are point-dependent quantities, while the energy of the photons themselves is unchanged along photons path's in a flat spacetime background.

%-------------------------------------------------

\subsubsection{Experimental evidence and geometry} 

The overwhelming situation discussed above is that, from the experimental point of view, extremely different yet physically equivalent descriptions, are perfectly allowed and equally ``real'' (here we are assuming that the EGR is a correct theory of gravity). Hence, the experimental evidence can not differentiate between the different representations within a conformal equivalence class. In other words, assuming that GR is a correct description of the laws of gravity, amounts to assuming that any of the infinitely many physically equivalent gauges: $(g^q_{\mu\nu},\vphi_q)$ ($q\geq 0$), $$g^q_{\mu\nu}=\Omega^2_q g^\text{GR}_{\mu\nu},\;\vphi_q=\vphi_0-2\ln\Omega_q,$$ where $\vphi_0$ is a constant, can be also a correct description of the laws of gravity. The experiment is not able to differentiate between these equivalent representations. In this regard the interesting question would not be, for instance, whether the expansion of the universe is accelerating or not, but, which one of the infinitely many physically equivalent conformal universes is the one we live in.

%%%%%%%%%%%%%%%%%%%%%%%%%%%%%%%%%%%%%%

\section{Conclusion}\label{sec-conclu}

The aim of this paper has been twofold. On the one hand our starting purpose was to discuss (once again) on the conformal transformations' issue. The study of conformal symmetry led us, on the other hand, to explore Weyl-invariant (also conformal or local scale invariant) theories of gravity -- including the SMP sector -- and their physical implications. In what regards to the physical equivalence of the conformal frames in which a given scalar-tensor theory of gravity (including the BD theory) may be formulated, our conclusion is that the different frames amount to different -- yet mathematically equivalent -- theories. Physical equivalence of the conformal frames requires invariance of the laws of gravity (and of the remaining laws of physics) under the Weyl rescalings \eqref{weyl-t}. This is not the case for the BD and the scalar-tensor theories, where the Weyl rescalings take us from one frame (say, the JF) into the conformal frame (the EF, for instance). Hence, the mere existence of the different frames is an evidence of the lack of conformal symmetry of these theories.

One of the most immediate implications of assuming that Weyl invariance is an actual symmetry of the laws of physics, is the gauge freedom that adds to the coordinate freedom inherent in the known fundamental theories. It happens that instead of a definite solution of the motion equations -- like, for instance, in GR -- a whole set of physically equivalent conformal solutions arises. All of these solutions are consistent with the same experimental data and none if preferred over the others. Hence, in a Weyl invariant world -- with general relativity as a particular gauge -- the question about whether the experiment can determine the curvature of the space, can not be consistently settled. If the EGR were a correct (classical) theory of gravity -- even after incorporating the SMP in a Weyl-invariant way (see subsection \ref{subsect-weyl-smp}) -- a space with curvature can be physically equivalent to a flat space with Weyl-integrable affinity, where the units of measure are point-dependent quantities. 

It may be that the question raised by Riemann some 164 years ago about the possibility to measure the curvature of space by means of physical experimentation, may not have a definitive answer after all.

%%%%%%%%%%%%%%%%%%%%%%%%%

%%%%%%%%%%%%%%%%%%%%%%%%%

\section{acknowledgment}

We thank J. Rubio, J. Hunter, L. J$\ddot a$rv, H. Veerm$\ddot a$e and C. Steinwachs for comments and for pointing us to several bibliographic references that are related to the present work. The authors acknowledge CONACyT of Mexico for continuous support. R De Arcia also acknowledges PRODEP by support of the present research.

%%%%%%%%%%%%%%%%%%%%%%%%%
%%%%%%%%%%%%%%%%%%%%%%%%%

\appendix

\section{Conformal transformations of the metric}\label{sect-conf-t}

Under the conformal transformation of the metric \eqref{conf-t} the affine connection coefficients (the Christoffel symbols) transform like:

\bea &&\{^\sigma_{\mu\nu}\}\rightarrow\{^\sigma_{\mu\nu}\}-\delta^\sigma_\mu\nabla_\nu\left(\ln\Omega\right)\nonumber\\
&&\;\;\;\;\;\;\;\;\;\;\;\;\;\;\;\;\;\;\;\;\;-\delta^\sigma_\nu\nabla_\mu\left(\ln\Omega\right)+g_{\mu\nu}\nabla^\sigma\left(\ln\Omega\right),\label{aff-conf-t}\eea while the components of the Ricci tensor and the curvature scalar transform in the following form:

\bea &&R_{\mu\nu}\rightarrow R_{\mu\nu}+2\nabla_\mu\left(\ln\Omega\right)\nabla_\nu\left(\ln\Omega\right)-2g_{\mu\nu}\left(\nabla\ln\Omega\right)^2\nonumber\\
&&\;\;\;\;\;\;\;\;\;\;\;\;\;\;\;\;\;\;\;\;\;\;\;\;\;\;\;\;\;\;\;+2\nabla_\mu\nabla_\nu\left(\ln\Omega\right)+g_{\mu\nu}\Box\ln\Omega,\nonumber\\
&&R\rightarrow\Omega^2\left[R+6\Box\ln\Omega-6\left(\nabla\ln\Omega\right)^2\right].\label{ricci-conf-t}\eea Taking into account the above transformation laws we can write the corresponding transformation law for the Einstein's tensor:

\bea &&G_{\mu\nu}\rightarrow G_{\mu\nu}+2\nabla_\mu\left(\ln\Omega\right)\nabla_\nu\left(\ln\Omega\right)+g_{\mu\nu}\left(\nabla\ln\Omega\right)^2\nonumber\\
&&\;\;\;\;\;\;\;\;\;\;\;\;\;\;\;\;\;\;\;\;\;\;\;\;\;\;\;+2\nabla_\mu\nabla_\nu\left(\ln\Omega\right)-2g_{\mu\nu}\Box\ln\Omega.\label{etensor-conf-t}\eea The transformation law for the D'Alembertian of a scalar field $\psi$ under \eqref{conf-t} reads:

\bea \Box\psi\rightarrow\Omega^2\left[\Box\psi-2\nabla^\sigma\left(\ln\Omega\right)\nabla_\sigma\psi\right].\label{box-conf-t}\eea

For the extrinsic curvature $K_{\mu\nu}=h^\sigma_\mu h^\lambda_\nu\nabla_\sigma n_\lambda$ ($K=h^{\mu\nu}K_{\mu\nu}$), of a 3D hypersurface ortogonal to the unit vector $n_\mu$ with the metric $h_{\mu\nu}=g_{\mu\nu}\pm n_\mu n_\nu$, induced on it, the transformation under \eqref{conf-t} reads:

\bea &&K_{\mu\nu}\rightarrow\Omega^{-1}\left(K_{\mu\nu}-h_{\mu\nu}n^\sigma\der_\sigma\Omega\right)\nonumber\\
&&\;\;\;\;\;\;\;\;\;\;\;\;\;\;\;\;\;\;\Rightarrow\;K\rightarrow\Omega^{-1}\left(K-3n^\sigma\der_\sigma\Omega\right).\label{ct-ext-curv}\eea

%--------------------------------------------------------------

\section{Quantum effects of matter}\label{sect-callan}

Here we give a brief account of the demonstration given in Ref. \cite{callan_ann_phys_1970}. Let us consider the simplest renormalizable quantum field theory that is given by the Lagrangian:

\bea {\cal L}_\phi=-\frac{1}{2}(\der\phi)^2-\frac{1}{2}\mu_0^2\phi^2-\lambda_0\phi^4.\label{renorm-lag}\eea The conventional stress-energy tensor $$T_{\mu\nu}^{(\phi)}=\der_\mu\phi\der_\nu\phi+g_{\mu\nu}{\cal L}_\phi,$$ does not have finite matrix elements, however, the modified tensor:

\bea \Theta_{\mu\nu}^{(\phi)}=T_{\mu\nu}^{(\phi)}+\frac{1}{6}\left(\nabla_\mu\nabla_\nu-g_{\mu\nu}\Box\right)\phi^2,\label{mod-set}\eea has finite matrix elements to all orders in $\lambda$. 

When we take into account the gravitational interactions, if we want the gravitational effects to be finite in to lowest order in the gravitational coupling and to all orders in all the other couplings, then, in the RHS of the Einstein's (GR) motion equations: $$G_{\mu\nu}=\frac{1}{M^2_\text{Pl}}T^{(\phi)}_{\mu\nu},$$ one has to make the replacement: $T^{(\phi)}_{\mu\nu}\rightarrow\Theta_{\mu\nu}^{(\phi)}$. This means, in turn, that the Einstein-Hilbert action principle: $$S_\text{EH}=\int d^4x\sqrt{|g|}\left[\frac{M^2_\text{Pl}}{2}R+{\cal L}_\phi\right],$$ is to be replaced by the STT action: $$S_*=\int d^4x\sqrt{|g|}\left[f(\phi)R+{\cal L}_\phi\right],$$ where $f(\phi)=M^2_\text{Pl}/2-\phi^2/12.$ 

Hence, in order to have observable gravitational effects when calculating, for instance, the amplitude of the scattering of a graviton in an external field, one has to rely on the stress-energy tensor that has finite matrix elements, i. e., on $\Theta_{\mu\nu}^{(\phi)}$. This, in turn, requires of a STT from the start.

%%%%%%%%%%%%%%%%%%%%%%%%%%%
%%%%%%%%%%%%%%%%%%%%%%%%%%%

\end{document}